\documentclass[twocolumn,amsmath,amssymb,floatfix,pre,unsortedaddress]{revtex4} 

\usepackage{graphicx,color}
\definecolor{brown}{rgb}{0.63,0.27,0.18}
\definecolor{orange}{rgb}{1.00,0.65,0.00}

\usepackage{dcolumn}
\usepackage{bm,multirow}

\begin{document}

\newcommand {\rsq}[1]{\langle R^2 (#1)\rangle}
\newcommand {\rsqL}{\langle R^2 (L) \rangle}
\newcommand {\rsqn}{\langle R^2 (n) \rangle}
\newcommand {\rsqbp}{\langle R^2 (N_{bp}) \rangle}
\newcommand {\Nbp}{N_{bp}}
\newcommand {\etal}{{\it et al.}}

\newcommand{\RalfNew}[1]{\textcolor{red}{#1}}



\title{Local loop opening in untangled ring polymer melts: \\ A detailed ``Feynman test'' of models for the large scale structure}

\author{Raoul D. Schram}
\email{raoul.schram@ens-lyon.fr}
\affiliation{
Univ Lyon, ENS de Lyon, Univ Claude Bernard, CNRS, Laboratoire de Physique and Centre Blaise Pascal, F-69342 Lyon, France}

\author{Angelo Rosa}
\email{anrosa@sissa.it}
\affiliation{
SISSA - Scuola Internazionale Superiore di Studi Avanzati, Via Bonomea 265, 34136 Trieste, Italy
}

\author{Ralf Everaers}
\email{ralf.everaers@ens-lyon.fr}
\affiliation{
Univ Lyon, ENS de Lyon, Univ Claude Bernard, CNRS, Laboratoire de Physique and Centre Blaise Pascal, F-69342 Lyon, France
}

\date{\today}

\begin{abstract}
The conformational statistics of ring polymers in melts or dense solutions is strongly affected by their quenched microscopic topological state. The effect is particularly strong for untangled ({\it i.e.} non-concatenated and unknotted) rings, which are known to crumple and segregate.
Here we study these systems using a computationally efficient multi-scale approach, where we combine massive simulations on the fiber level with the explicit construction of untangled ring melt configurations based on theoretical ideas for their large scale structure.
We find
(i)
that topological constraints may be neglected on scales below the standard entanglement length, $L_e$, 
(ii)
that rings with a size $1 \le L_r/L_e \le 30$ exhibit nearly ideal lattice tree behavior characterized by primitive paths which are randomly branched on the entanglement scale,
and
(iii)
that larger rings are compact with gyration radii $\langle R_g^2(L_r) \rangle \propto L_r^{2/3}$.
The detailed comparison between equilibrated and constructed ensembles allows us to perform a ``Feynman test'' of our understanding of untangled rings:
can we convert ideas for the large-scale ring structure into algorithms for constructing (nearly) equilibrated ring melt samples? 
We show that most structural observables are quantitatively reproduced by two different construction schemes: hierarchical crumpling and ring melts derived from the analogy to interacting branched polymers.
However, the latter fail the ``Feynman test'' with respect to the magnetic radius, $R_m$, which we have defined based on an analogy to magnetostatics.
While $R_m$ is expected to vanish for double-folded structures, the observed values of $\langle R_m^2(L_r) \rangle \propto \langle R_g^2(L_r) \rangle$ provide a simple and computationally convenient measure of the presence of a non-negligible amount of local loop opening in crumpled rings.
\end{abstract}

\maketitle

\section{Introduction}\label{sec:intro}
Similar to macroscopic strings tied into knots, the (Brownian) motion of polymer chains is subject to topological constraints:
they can slide past each other, but their backbones cannot cross~\cite{Edwards_procphyssoc_67,PragerFrisch_67}.
For {\it linear} chains, the constraints are transient and irrelevant for the {\it equilibrium} statistics:
chains with a contour length exceeding the material specific Kuhn length, $L \gg l_K$,
show Gaussian behavior with mean-square end-to-end distances $\langle R^2(L) \rangle=l_K L$.
The only effect of the constraints is to slow down the chain dynamics beyond a density dependent entanglement (contour) length, $L_e$,
a corresponding spatial distance or ``tube'' diameter, $d_T = \sqrt{l_K L_e / 6}$~\cite{LooselyEntangledNote},
and a characteristic entanglement time, $\tau_e$~\cite{DoiEdwards,mcleish2002}.

The situation is different for {\it un}tangled polymer melts or solutions,
where the chain conformations have to respect (long-lived) {\it global} constraints enforcing the {\it absence} of topological knots and links~\cite{rolfsen}.
Experimentally prepared systems of this type have interesting materials properties~\cite{Spiess2005,kapnistos2008}.
With large (interphase) chromosomes~\cite{grosbergEPL1993,RosaPLOS2008,hic,Vettorel2009,Grosberg_PolSciC_2012,DiStefanoRosa2013,DekkerMartiRenomMirny2013}
the most prominent representatives are probably found in biological systems. 
In this case, the relaxation times for the topological state may be of the order of centuries~\cite{SikoravJannink,RosaPLOS2008},
making the entanglement-free state sufficiently long lived to merit attention. 

The prototype untangled polymer liquid is a melt of non-concatenated unknotted ring polymers~\cite{KhokhlovNechaev85,CatesDeutsch,klein_ring,RubinsteinPRL1986,BreretonVilgis1995,mullerPRE1996,mullerPRE2000,Vettorel2009,DeguchiJCP2009,Halverson2011_1,HalversonPRL2012,Grosberg_PolSciC_2012}. 
As a measure of how difficult it is to understand these systems consider the spread in the proposed values for the characteristic exponent, $\nu$, which relates mean-square gyration radius and contour length, $\langle R_g^2(L_r) \rangle \propto L_r^{2\nu}$. 
Plausible values range from
$\nu=1/4$
for ideal lattice trees or animals~\cite{KhokhlovNechaev85,ORD_PRL1994},
$\nu=1/3$
for crumpled~\cite{grosbergJPhysFrance1988} or ``loopy'' globules~\cite{ObukhovWittmerEPL2014,PanyukovRubinsteinMacromolecules2016}, Hamiltonian paths~\cite{hic,SmrekGrosberg2013} and interacting lattice trees~\cite{KhokhlovNechaev85,GrosbergSoftMatter2014},
$\nu=2/5$~\cite{CatesDeutsch} from a Flory argument balancing the entropic cost of compressing Gaussian rings and the unfavorable overlap with other chains
(recently refined to $\nu=1/3$ for the asymptotic behavior~\cite{SakauePRL2012}), to
$\nu=(1-1/(3\pi))/2\approx0.45$~\cite{BreretonVilgis1995}, and
$\nu=1/2$ for Gaussian rings, rings folded into linear ribbons~\cite{klein_ring} and swollen lattice trees~\cite{RubinsteinPRL1986}.
Thirty years after the pioneering theoretical studies, there is now strong numerical evidence~\cite{mullerPRE1996,mullerPRE2000,Vettorel2009,DeguchiJCP2009,Halverson2011_1,HalversonPRL2012} that untangled rings  exhibit a crossover to marginally overlapping, ``crumpled'' configurations with $\nu=1/3$ for ring sizes around $Z_r \equiv L_r / L_e = 10$~\cite{HalversonPRL2012}.
However, this by itself is not sufficient to decide which (if any) of the available compatible models truly describes their structure.

In Ref.~\cite{hic}, Lieberman-Aiden {\it et al.} have investigated a large number of unknotted,
fractal space-filling curves from the mathematical literature as simplified representations of crumpled or fractal globules~\cite{grosbergJPhysFrance1988,grosbergEPL1993,hic},
while in Ref.~\cite{SmrekGrosberg2013} Smrek and Grosberg presented whole new sets of recursive fractal curves characterized by non-trivial fractal-like surfaces.
As an alternative, Tamm {\it et al.}~\cite{Tamm-PRL2015} investigated surface-enhanced random walks.
In a recent letter~\cite{RosaEveraersPRL2014}, two of us extended this approach to a multi-scale ``Feynman test'' of our understanding of untangled rings: can we convert ideas for the large-scale ring structure into algorithms for constructing (nearly) equilibrated ring melt samples?  
The idea is (i) to generate coarse-grain ``Klein ribbons'', ``Moore rings'',  ``Hilbert rings'', or to fold rings around the outline of ideal or interacting lattice trees, (ii) to ``fine-grain'' them to the scale of the polymer models used for the ``gold standard'' simulations, and (iii) to equilibrate the systems on the entanglement scale (see Fig. 1 in Ref.~\cite{RosaEveraersPRL2014}). 
The explicit construction allows to derive detailed predictions for averages and distribution functions of arbitrary structural observables, which can then be compared to ``gold standard'' reference data from brute-force equilibrated samples. This analysis extends far beyond the comparison of exponents for the asymptotic regime, which the ``gold standard'' simulations may or may not have reached.

We emphasize that success in this ``Feynman test'' is (i) relative and (ii) dependent on the observables used for the comparison. 
Significant deviations in a single observable are sufficient to establish the failure of a construction algorithm, while there is the obvious caveat that no number of successful comparisons can positively affirm the equivalence between a model-derived and the properly equilibrated ensemble.  
Within the limits of the approach we pursue two objectives. Firstly we aim to identify the physics underlying the crumpling of rings, secondly we seek to validate a multi-scale algorithm for generating plausible melt structures for otherwise inaccessible ring sizes. The use of such algorithms is common practice in the case of linear polymer melts, where corresponding algorithms~\cite{AuhlJCP2003,ZhangKremer2014,SvaneborgEveraers2016} exploit the well-understood large random walk statistics of ideal chains. 
But even the ability to generate a wide range of qualitatively different and hence untypical initial states has its uses, as it allows to validate the proper equilibration of our MD simulations at the fiber level~\cite{Vettorel2009,RosaEveraersPRL2014}. 

Reference~\cite{RosaEveraersPRL2014} concluded that the large-scale behavior of untangled ring melts is well reproduced by the model of interacting lattice trees, which was proposed in some of the very first ring melt studies~\cite{KhokhlovNechaev85,RubinsteinPRL1986}. In this view, crumpling can be understood by the successive application of three different strategies for entropy maximization: double-folding, branching, and swelling.
Firstly, and most importantly, the rings adopt double-folded configurations to minimize the threadable surface as this reduces the importance of the topological constraints they impose on each other.
The simplest example for this strategy are melts of linear ``Klein"~\cite{klein_ring} ribbons, which are double-folded on the tube scale and which can freely pack into dense melt configurations. 
Secondly, and in contrast to linear chains, double-folded rings can increase their entropy by branching.
This effect is clearly visible for $Z_r > {\cal O}(10)$. Systems exhibiting random branching being asymptotically too dense, excluded volume interactions are, thirdly, not fully screened. Lattice tree melts hence exhibit a small amount of swelling with a slightly modified branching statistics~\cite{GrosbergSoftMatter2014,RosaEveraersJCP2016}. These effects becomes relevant for $Z_r > {\cal O}(100)$.

But can this really be all? If most of the contour length is neatly tugged away in crumpled, double-folded ring sections, then rings can safely open {\it some} loops {\it without} violating topological constraints~\cite{ObukhovWittmerEPL2014} as long as their concentration is below the overlap concentration at the respective scale~\cite{PanyukovRubinsteinMacromolecules2016}. 
Such openings are indeed apparent in simulation snapshots~\cite{Halverson2011_1,Halverson2011_2} and first studies of the minimal surface enclosed by small and medium sized crumpled rings~\cite{LangMacromolecules2013,SmrekGrosbergACSMacroLett2016} 
reveal a limited amount of mutual threading at equilibrium. 
It remains to be seen if rare deep threading of open loops leads to a topological glass transition in the limit of infinite chain length~\cite{TsalikisVlassopoulos2016,MichielettoTurnerPNAS2016,MichielettoRosaPRL2017}.
However, it is now clear~\cite{SmrekKremerRosaACSML2019} that the interacting lattice tree derived untangled melts from Ref.~\cite{RosaEveraersPRL2014} fail the ``Feynman test'' with respect to these new observables.
As a consequence, we are faced with the challenge to devise a different construction algorithm for untangled ring melts, which
(i) passes a more complete ``Feynman test''
and
(ii) allow to study larger ring sizes than those accessible to brute-force equilibration.

The present article significantly extends our multi-scale exploration of the structure of untangled polymers.
Besides providing a more complete account of the data presented in Ref.~\cite{RosaEveraersPRL2014},
(i) we add additional ``gold standard'' Monte Carlo results for a highly efficient lattice model from Refs.~\cite{SchramBarkemaSchiessel2013,SchramBarkema2018},
(ii) we construct melt configurations for large untangled rings through a ``hierarchical crumpling'' algorithm for the lattice model, and
(iii) we introduce and analyze with the ``magnetic moment'' vector and the corresponding ``magnetic radius'' two easily computable observables designed to detect the presence of significant open loops in crumpled ring structures.

The paper is organized as follows:
In Sec.~\ref{sec:Methods}, we summarize the numerical models and methods as well as the units of length and time employed in this work.
Additional details concerning the construction of melts of rings with different spatial conformations and their road to equilibrium are provided in the Supplemental Material (SM).
The reader aiming to more advanced discussion may look into former works~\cite{RosaEveraersPRL2014,lammps,SchramBarkema2018}.
The main results of this work are presented and discussed in Sections~\ref{sec:Results} and~\ref{sec:Discussion} respectively,
while in Sec.~\ref{sec:Concls} we outline the conclusions.

\section{Models and Methods}\label{sec:Methods}
The two polymer models and the simulation codes we have employed for the acquisition of the reference data have been described elsewhere~\cite{RosaEveraersPRL2014,lammps,SchramBarkema2018} in sufficient detail, so that we only present short summaries in Sections~\ref{sec:fiberModel} and~\ref{sec:Elastic lattice polymer model}. 
The entanglement units, which we employ throughout the article, are explained in Sec.~\ref{sec:Entanglement units}. 

We distinguish two classes of  techniques for generating theoretically inspired ensembles of untangled ring melts. 
The first class, described in Sec.~\ref{sec:Lattice model of double folded rings}, imposes double folding and employs Monte Carlo techniques to simulate lattice polymer models for the linear or branched primitive path. 
These primitive paths are subsequently ``wrapped'' by a tightly double-folded bead-spring chain, assembled into solution conformations and then locally equilibrated over the entanglement time. 
Again we limit our account to a short summary. Detailed information can be found in Refs.~\cite{RosaEveraersJPhysA2016,RosaEveraersJCP2016,RosaEveraersPRE2017,RosaEveraersDoubleFolding2019}.
The computational strategy to pass from coarse-grain to the fiber model is described in detail in the Supplemental Material of Ref.~\cite{RosaEveraersPRL2014}. 
The second class, described in more detail in Sec.~\ref{sec:Fractals},
employs fractal building algorithms for space-filling curves. In particular, we have combined this idea with the lattice polymer model of ring polymers into a ``hierarchical crumpling'' algorithm.

\subsection{Off-lattice fiber model and Molecular Dynamics simulations}\label{sec:fiberModel}
We used a variant~\cite{RosaPLOS2008} of the Kremer-Grest~\cite{KremerGrestJCP1990} bead-spring polymer model to study ring polymers at the fiber level.
The model accounts for the connectivity, bending rigidity, excluded volume
and topology conservation of polymer chains.
Specifically, beads of diameter $\sigma$ interact via a purely repulsive Weeks-Chandler-Andersen potential and are connected into rings by finite-extensible-nonlinear-elastic (FENE) springs. Due to a weak bending potential, the chains have a Kuhn length of $l_K = 10.0\sigma$. With a bead density of $\rho=0.1\sigma^{-3}$ our systems are relatively dilute. Chain dynamics was studied by using fixed-volume Molecular Dynamics simulations,
with Langevin thermostat in order to keep fixed the temperature of the system.
The system dynamics was integrated by using LAMMPS~\cite{lammps}.
Details on the initial states and the total computational effort are given in Sec.~\ref{sec:Lattice model of double folded rings} here, Sec.~\ref{sec:SM Fractals} in SM and Ref.~\cite{RosaEveraersPRL2014}.
While we have not performed additional simulations compared to Ref.~\cite{RosaEveraersPRL2014},
we have added material to the SM (Secs.~\ref{sec:SupplMat:RingsStructure} and~\ref{sec:Brute force equilibration}) of the present article to better document the equilibration of our ``gold standard'' MD results.

\subsection{Elastic lattice polymer model and Monte Carlo simulations}\label{sec:Elastic lattice polymer model}
As a complement, we employ the numerically much more efficient~\cite{SchramBarkema2018} Monte Carlo simulations of an ``elastic''~\cite{BarkemaElasticModel2005} lattice polymer model.
The model resides on a FCC lattice, multiple occupation of lattice sites is limited to consecutive monomers belonging to the same chain.
Here we use a GPU algorithm limited to maximal occupancy equal to 2.
A bond with length $=0$ as a result of two monomers occupying the same site is a unit of stored length. Monte Carlo moves are divided into two categories: moves that displace stored length along the backbone of the polymer and moves that transform a unit of stored length and a normal bond into two regular bonds and back.
Rings composed of $N$ monomers have a contour length of $L_r \approx 0.714N b$ in units of the bond length, $b$, of the FCC lattice.
They are highly flexible with Kuhn length $l_K\approx1.47 b$.
The number of Kuhn segments per ring is thus given by $N_K = L_r / l_K \approx 0.486 N$. 
A more detailed description of the algorithm is available in~\cite{SchramBarkema2018}.

\subsection{Entanglement units}\label{sec:Entanglement units}
The relevant length and time scales for the topological effects we are interested in are, respectively, the entanglement length ($L_e \equiv l_K \, N_{eK}$) and the entanglement time ($\tau_e$)~\cite{HalversonPRL2012}.

For our polymer models, $N_{eK}$ can be estimated from the packing argument by Lin~\cite{LinMacromolecules1987} and by Kavassalis and Noolandi~\cite{KavassalisNoolandiPRL1987}.
The number of entanglement strands sharing the volume spanned by one entanglement strand,
\begin{equation}\label{eq:KavassalisNoolandi}
\frac{\rho_K}{N_{eK}} \langle R^2(N_{eK}) \rangle^{3/2} \approx 20 \, ,
\end{equation}
appears to be a universal constant for all flexible polymers~\cite{LinMacromolecules1987,KavassalisNoolandiPRL1987,FettersMacromolecules1994,RosaEveraersPRL2014}, suggesting $N_{eK} \approx (20 / \rho_K l_K^3)^2$.
For the MD model $\rho_K l_K^3 = 10$ implies $N_{eK} \approx 4$ or $N_e=40$ monomers.
For the lattice model, $\rho_K l_K^3 = 2.62$ implies $N_{eK} \approx 59$ or $N_e=121$ monomers.
The corresponding unit of distance is the tube diameter, $d_T^2 \equiv \langle R_g^2(N_e) \rangle = l_K L_e/6$.
For our two models, we find:
$d_T^2 \approx 0.67 \, l_K^2 = 66.67 \sigma^2$ (MD model)
and
$d_T^2 \approx 9.83 \, l_K^2 \approx 21.25b^2$ (lattice model).

Similarly, the corresponding entanglement time, $\tau_e$, can be defined~\cite{KremerGrestJCP1990} as the time when the monomer mean-square displacement reaches the tube diameter, $g_1(\tau_e) \equiv 2 d_T^2$. 
For the MD model, $\tau_e \approx 1.6\cdot10^3\tau_{LJ}$~\cite{RosaPLOS2008} where $\tau_{LJ} = \sigma \sqrt{m / \epsilon}$ is the elementary Lennard-Jones (LJ) time unit of the simulation protocol expressed as a function of units of length ($\sigma$), mass ($m$) and energy ($\epsilon$).
For the MC model, $\tau_e \approx 5\cdot10^4 \tau_{MC}$ where $\tau_{MC}$ is the elementary Monte Carlo time unit.

Typically, we will present our results in these units to simplify the comparison between the two models.

Beyond the entanglement scale, the behavior of the two polymer models is expected to agree and to reproduce the universal behavior of loosely entangled polymers.
On smaller scales, differences can be expected.
With $N_{eK}=59$ the entanglement length of the lattice model is significantly larger than the Kuhn length. As a consequence, the chains exhibit flexible chain behavior on the entanglement scale. 
In contrast, there is no pronounced Rouse regime in the off-lattice model with $N_{eK}=4$.
Mapping from the elastic lattice polymer model to the fiber model is, in principle, possible through a procedure resembling the primitive path analysis~\cite{everaers_science}, which reduces the contour length and increases the effective stiffness while preserving the microscopic topological state of the samples.

\subsection{Lattice models and Monte Carlo simulations of the linear or branched primitive paths of double-folded ring polymers}\label{sec:Lattice model of double folded rings}
We have modelled the primitive paths characterizing double-folded ring conformations on a simple cubic lattice with lattice constant $l_K$.
Each primitive path segment represents two ring segments, the primitive path contour density is thus half the ring contour density, $(\rho_{K}^{pp} l_K^3)=5$.

As pointed out by Klein~\cite{klein_ring}, 
it is not immediately obvious that non-concatenated rings fold into compact conformations when brought into contact:
rings, which double-fold on the entanglement scale and adopt linear ribbon conformations can retain substantial conformational entropy, while threading between topological obstacles in the same way as a linear chain.
By choosing~\cite{RosaEveraersPRL2014} the same Kuhn length for the ribbon axis as for the fiber model,
a straightforward Monte Carlo procedure allows to generate corresponding linear primitive paths in the form of $(L_r/l_K)/2$-step random walks. 
A better model~\cite{KhokhlovNechaev85} are ideal lattice trees, where a comparison to our MD simulations of the fiber model suggests a branching probability of $\lambda \approx 0.4 / l_K$ \cite{RosaEveraersPRL2014}. 
To generate ideal lattice trees we have used the ``amoeba'' algorithm of Seitz and Klein~\cite{seitz_klein}, which is a straightforward generalization of the reptation scheme for linear chains.
For details on the simulation procedure and for our results on their conformational statistics we refer the reader to Refs.~\cite{RosaEveraersJPhysA2016,RosaEveraersPRE2017}. 

\begin{table}
\begin{tabular}{ccccccc}
$Z_r$ & $N \times M \times \#\mbox{RUNS}$ & $\tau_{tot} [\times 10^4]$ & $\tau_{tot} / \tau_{eq}$ \\
\hline
\hline
1.5 & $3\times160\times100$ & $1$ & $\approx1000$ \\
2.5 & $5\times64\times100$ & $1$ & $\approx200$ \\
5 & $10\times32\times100$ & $1$ & $\approx20$ \\
15 & $30\times256\times100$ & $2$ & $\approx3$ \\
37.5 & $75\times256\times25$ & $18$ & $\approx2$ \\
115 & $230\times256\times25$ & $430$ & $\approx2$ \\
225 & $450\times256\times25$ & $943$ & $\approx2$ \\
450 & $900\times128\times25$ & $4335$ & $\approx2$ \\
900 & $1800\times64\times25$ & $19922$ & $\approx2$ \\
\hline
\end{tabular}
\caption{
\label{tab:MCruns}
Monte Carlo simulations of lattice tree (LT) melts (Sec.~\ref{sec:Lattice model of double folded rings}).
$Z_r$: number of entanglements per each LT;
$N$: total mass of each LT in number of lattice units;
$M$: total number of independent chains per each melt configuration;
$\#\mbox{RUNS}$: total number of independent MC trajectories;
$\tau_{tot}$: total number of MC steps per single polymer;
$\tau_{tot} / \tau_{eq}$: total number of independent MC configurations,
$\tau_{eq}$ is the equilibration time estimated by comparing the mean-square displacement of the ring center of mass to the ring square gyration radius
(see Fig.~S3b in Supplemental Material of Ref.~\cite{RosaEveraersPRL2014}).
}
\end{table}

A realistic model has to account for the partially screened excluded volume interactions in a dense solution~\cite{DaoudJoanny1981,KhokhlovNechaev85,GrosbergSoftMatter2014,EveraersRosaFloryReview2017}.
Our simulations of lattice-tree melts are described in detail in Refs.~\cite{RosaEveraersJCP2016,RosaEveraersPRE2017}.
Compared to brute-force Molecular Dynamics equilibration the speedup is of the order of $\approx 10^6$ for our largest ring sizes~\cite{RosaEveraersPRL2014}.
A summary of the studied systems is given in Table~\ref{tab:MCruns}.

The ``wrapping'' of ring polymers around the primitive trees and the conformational statistics of the resulting ring ensembles is discussed in Ref.~\cite{RosaEveraersDoubleFolding2019}.
Finally, we refer the reader to the SM of Ref.~\cite{RosaEveraersPRL2014} for details on the fine-graining procedure to the off-lattice fiber model.
Fig.~\ref{fig:hierarchical} (top row) illustrates a few ring conformations from $Z_r\approx15$ up to $Z_r\approx900$.

\subsection{Building ring melts using algorithms for constructing fractals}\label{sec:Fractals}
A completely different class of models for untangled rings derives from the analogy to crumpled or fractal globules~\cite{grosbergJPhysFrance1988,grosbergEPL1993,hic,SmrekGrosberg2013}.
Taken literally, the analogy suggests to build ring melts from compact single chain conformations
generated by a rapid mechanical confinement~\cite{hic} or a non-equilibrium collapse~\cite{AbramsEPL2002,SchramBarkemaSchiessel2013}
of chains with random walk or self-avoiding walk statistics.  
In Ref.~\cite{RosaEveraersPRL2014} we have used the classical Hilbert curve~\cite{Hilbert1891,mandelbrot} and its closed variant, the Moore curve~\cite{SpaceFillingCurvesSagan,SpaceFillingCurvesVentrella}, to construct ring melts composed of non-overlapping space-filling rings.
Details on the construction procedure for these systems can be found in Sec.~\ref{sec:SM Fractals} in SM. 

Here we have used a similar approach to devise a ``hierarchical crumpling'' (HC) scheme for the elastic lattice polymer model (Sec.~\ref{sec:Elastic lattice polymer model}).  
Starting from ``gold standard'' equilibrium ring melt conformations, we construct larger systems in two steps~\cite{SchramBarkemaSchiessel2013}:

{\it Step 1} --
In analogy to the refinement of Hilbert or Moore curves from one iteration to the next, we refine
the linear dimensions of the lattice by a factor of two and split the original monomers into eight to keep the monomer density constant. 
In the original version of the elastic lattice polymer~\cite{BarkemaElasticModel2005}, it is trivial to find suitable positions for the new monomers:
they can simply be placed in equal numbers at and between the positions of the original monomers keeping the chain contour unchanged.
Occupied sites are then initially filled by multiples of four monomers with the density quickly equilibrating after a few MC steps.
As we are building on the results from Ref.~\cite{SchramBarkema2018}, we have continued to employ the GPU version of the algorithm, which does not allow for site occupancies exceeding two. 
As a consequence, the distribution of the new monomers becomes more complicated.
A point to note is that we can exclusively assign cells containing ``$2\times2\times2$'' sites on the new lattice to one site of the old lattice. 
As a consequence, it is possible to find arrangements where each of the new sites is occupied by as many new monomers as there were monomers present on the original site ({\it i.e.} zero, one or two). 
What remains to be done is to define the order in which the new monomers are connected. To do so, for each original bond, we have to place one corresponding inter-cell bond. Furthermore, for each original monomer we have to find a linear sequence of 7 intra-cell bonds linking the entry and exit points of the chain through all the monomers within the cell.
On the FCC lattice, this is possible for arbitrary entry and exit points.
Since the entry and exit points of a cell have to be distinct, some care has to be taken in selecting them among the $1$, $2$ or $4$ different possibilities for linking cells with neighbors with which they share a point, a line or a surface, respectively.
Moreover the list of possibilities needs to be continuously updated during the construction of the chain. 
We encountered no problems using a greedy algorithm, starting from the most constrained bonds and choosing randomly when presented with multiple possibilities.
As soon as all the entry and exit points are known, we define the intra-cell bonds cell by cell choosing randomly between the available options. 

\begin{figure}
\includegraphics[width=0.48\textwidth]{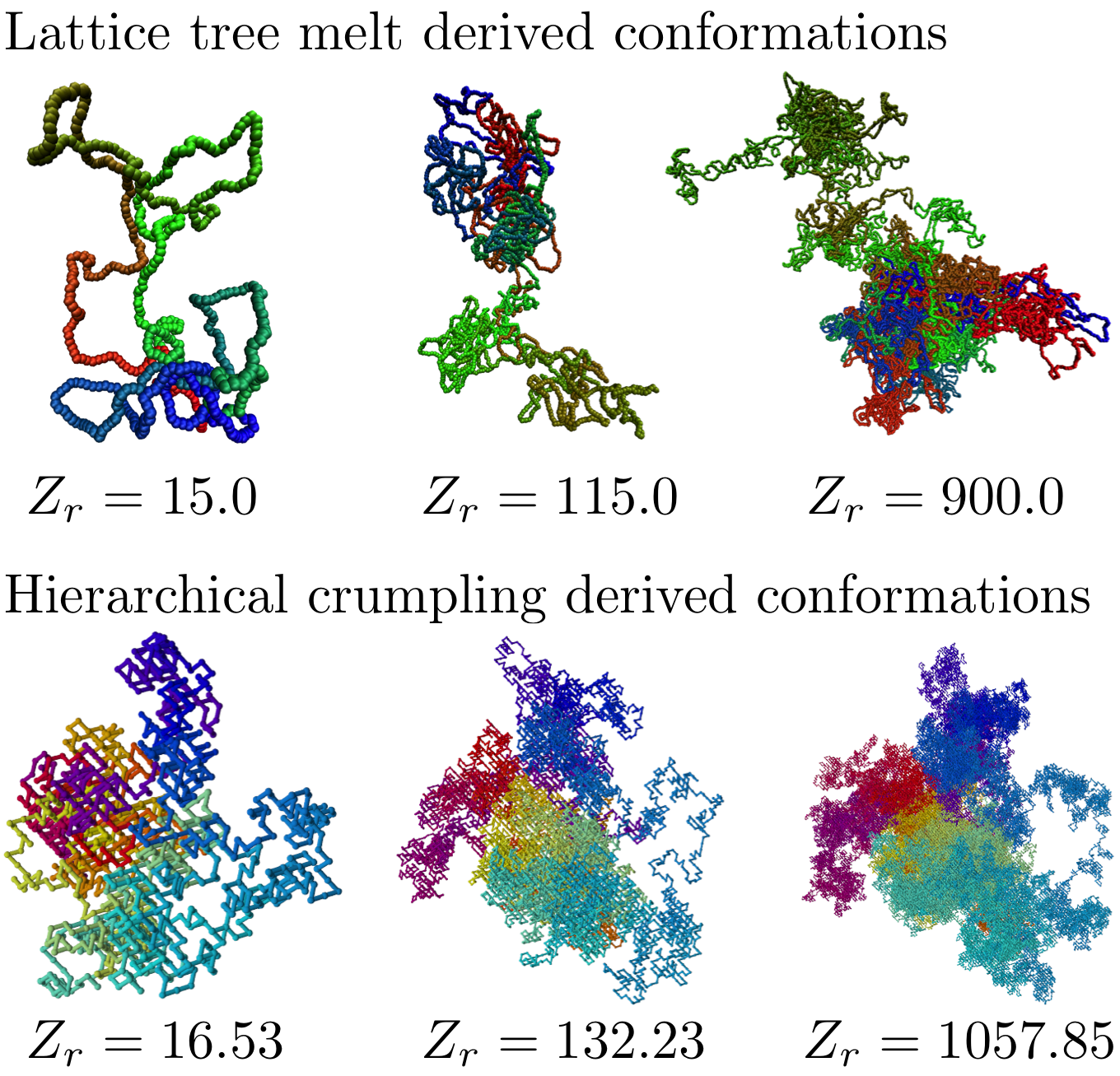}
\caption{
\label{fig:hierarchical}
(Top row)
Single ring conformations of increasing (from left to right) contour length $Z_r$ based on the lattice tree melt derived ensemble (Sec.~\ref{sec:Lattice model of double folded rings}).
Each melt contains a total number of chains from $256$ to $64$ (see Table~\ref{tab:MCruns}).
(Bottom row)
Single ring conformations from crumpled ring melts based on the hierarchical building algorithm (Sec.~\ref{sec:Fractals}).
From left to right are shown the first, second and third generation of a single ring polymer from such melts.
The first generation is an equilibrated melt of $1258$ ring polymers with $Z_r \approx 16$, increasing to $Z_r \approx 1000$ for the last generation.
The largest ring polymers studied in this paper with this construction method are twice as big as the one shown in the figure (see Table~\ref{tab:RaoulMCSimuls}). 
}
\end{figure}
\begin{table}
\begin{tabular}{cccccc}
$Z_r$ & $N \times M$ & $\tau_{tot} [\tau_{MC}]$ & $\tau_{eq} [\tau_{MC}]$ & $\tau_{tot}/\tau_{eq}$ & $M \cdot \tau_{tot}/\tau_{eq}$\\
\hline
\hline
{\footnotesize 0.25 } &  {\footnotesize $30\times 83886$} & {\footnotesize $1.0\cdot10^5$} & {\footnotesize $8.4\cdot10^3$} & {\footnotesize $1.2\cdot10^1$} & {\footnotesize $1.0\cdot10^6$}\\
{\footnotesize 0.50 } &  {\footnotesize $60\times 41943$} & {\footnotesize $5.2\cdot10^5$} & {\footnotesize $8.4\cdot10^3$} & {\footnotesize $6.2\cdot10^1$} & {\footnotesize $2.6\cdot10^6$}\\
{\footnotesize 0.83 } &  {\footnotesize $100\times 25166$} & {\footnotesize $1.0\cdot10^6$} & {\footnotesize $9.9\cdot10^3$} & {\footnotesize $1.0\cdot10^2$} & {\footnotesize $2.5\cdot10^6$}\\
{\footnotesize 1.24 } &  {\footnotesize $150\times 16777$} & {\footnotesize $3.0\cdot10^6$} & {\footnotesize $2.4\cdot10^4$} & {\footnotesize $1.3\cdot10^2$} & {\footnotesize $2.2\cdot10^6$}\\
{\footnotesize 1.65 } &  {\footnotesize $200\times 12583$} & {\footnotesize $5.0\cdot10^6$} & {\footnotesize $4.6\cdot10^4$} & {\footnotesize $1.1\cdot10^2$} & {\footnotesize $1.4\cdot10^6$}\\
{\footnotesize 2.48 } &  {\footnotesize $300\times 8389$} & {\footnotesize $3.0\cdot10^7$} & {\footnotesize $1.1\cdot10^5$} & {\footnotesize $2.7\cdot10^2$} & {\footnotesize $2.3\cdot10^6$}\\
{\footnotesize 3.10 } &  {\footnotesize $375\times 6711$} & {\footnotesize $5.0\cdot10^7$} & {\footnotesize $1.8\cdot10^5$} & {\footnotesize $2.8\cdot10^2$} & {\footnotesize $1.9\cdot10^6$}\\
{\footnotesize 4.13 } &  {\footnotesize $500\times 5033$} & {\footnotesize $7.0\cdot10^7$} & {\footnotesize $3.4\cdot10^5$} & {\footnotesize $2.0\cdot10^2$} & {\footnotesize $1.0\cdot10^6$}\\
{\footnotesize 5.79 } &  {\footnotesize $700\times 3595$} & {\footnotesize $2.0\cdot10^8$} & {\footnotesize $7.2\cdot10^5$} & {\footnotesize $2.8\cdot10^2$} & {\footnotesize $1.0\cdot10^6$}\\
{\footnotesize 8.26 } &  {\footnotesize $1000\times 2517$} & {\footnotesize $3.0\cdot10^8$} & {\footnotesize $1.9\cdot10^6$} & {\footnotesize $1.6\cdot10^2$} & {\footnotesize $4.0\cdot10^5$}\\
{\footnotesize 12.40 } &  {\footnotesize $1500\times 1678$} & {\footnotesize $5.0\cdot10^8$} & {\footnotesize $6.6\cdot10^6$} & {\footnotesize $7.6\cdot10^1$} & {\footnotesize $1.3\cdot10^5$}\\
{\footnotesize 16.53 } &  {\footnotesize $2000\times 1258$} & {\footnotesize $1.0\cdot10^9$} & {\footnotesize $1.6\cdot10^7$} & {\footnotesize $6.5\cdot10^1$} & {\footnotesize $8.2\cdot10^4$}\\
{\footnotesize 24.79 } &  {\footnotesize $3000\times 839$} & {\footnotesize $2.0\cdot10^9$} & {\footnotesize $5.5\cdot10^7$} & {\footnotesize $3.6\cdot10^1$} & {\footnotesize $3.0\cdot10^4$}\\
{\footnotesize 33.06 } &  {\footnotesize $4000\times 629$} & {\footnotesize $3.0\cdot10^9$} & {\footnotesize $1.3\cdot10^8$} & {\footnotesize $2.2\cdot10^1$} & {\footnotesize $1.4\cdot10^4$}\\
\hline
{\footnotesize 9.92 } &  {\footnotesize $1200\times 16777$} & {\footnotesize $5.0\cdot10^6$} & {\footnotesize $3.3\cdot10^6$} & {\footnotesize $1.5\cdot10^0$} & {\footnotesize $2.5\cdot10^4$}\\
{\footnotesize 13.22 } &  {\footnotesize $1600\times 12583$} & {\footnotesize $5.0\cdot10^6$} & {\footnotesize $8.0\cdot10^6$} & {\footnotesize $6.3\cdot10^{-1}$} & {\footnotesize $7.9\cdot10^3$}\\
{\footnotesize 19.83 } &  {\footnotesize $2400\times 8389$} & {\footnotesize $5.0\cdot10^6$} & {\footnotesize $2.8\cdot10^7$} & {\footnotesize $1.8\cdot10^{-1}$} & {\footnotesize $1.5\cdot10^3$}\\
{\footnotesize 24.79 } &  {\footnotesize $3000\times 6711$} & {\footnotesize $5.0\cdot10^7$} & {\footnotesize $5.5\cdot10^7$} & {\footnotesize $9.0\cdot10^{-1}$} & {\footnotesize $6.0\cdot10^3$}\\
{\footnotesize 33.06 } &  {\footnotesize $4000\times 5033$} & {\footnotesize $5.0\cdot10^6$} & {\footnotesize $1.3\cdot10^8$} & {\footnotesize $3.7\cdot10^{-2}$} & {\footnotesize $1.9\cdot10^2$}\\
{\footnotesize 46.28 } &  {\footnotesize $5600\times 3595$} & {\footnotesize $5.0\cdot10^6$} & {\footnotesize $3.8\cdot10^8$} & {\footnotesize $1.3\cdot10^{-2}$} & {\footnotesize $4.7\cdot10^1$}\\
{\footnotesize 66.12 } &  {\footnotesize $8000\times 2517$} & {\footnotesize $5.0\cdot10^6$} & {\footnotesize $1.1\cdot10^9$} & {\footnotesize $4.4\cdot10^{-3}$} & {\footnotesize $1.1\cdot10^1$}\\
{\footnotesize 99.17 } &  {\footnotesize $12000\times 1678$} & {\footnotesize $5.0\cdot10^6$} & {\footnotesize $4.0\cdot10^9$} & {\footnotesize $1.3\cdot10^{-3}$} & {\footnotesize $2.2\cdot10^0$}\\
{\footnotesize 132.23 } &  {\footnotesize $16000\times 1258$} & {\footnotesize $5.0\cdot10^6$} & {\footnotesize $9.6\cdot10^9$} & {\footnotesize $5.2\cdot10^{-4}$} & {\footnotesize $6.5\cdot10^{-1}$}\\
{\footnotesize 198.35 } &  {\footnotesize $24000\times 839$} & {\footnotesize $5.0\cdot10^6$} & {\footnotesize $3.4\cdot10^{10}$} & {\footnotesize $1.5\cdot10^{-4}$} & {\footnotesize $1.3\cdot10^{-1}$}\\
{\footnotesize 264.46 } &  {\footnotesize $32000\times 629$} & {\footnotesize $5.0\cdot10^6$} & {\footnotesize $8.1\cdot10^{10}$} & {\footnotesize $6.2\cdot10^{-5}$} & {\footnotesize $3.9\cdot10^{-2}$}\\
\hline
{\footnotesize 79.34 } &  {\footnotesize $9600\times 16777$} & {\footnotesize $5.0\cdot10^6$} & {\footnotesize $2.0\cdot10^9$} & {\footnotesize $2.5\cdot10^{-3}$} & {\footnotesize $4.2\cdot10^1$}\\
{\footnotesize 105.79 } &  {\footnotesize $12800\times 12583$} & {\footnotesize $5.0\cdot10^6$} & {\footnotesize $4.8\cdot10^9$} & {\footnotesize $1.0\cdot10^{-3}$} & {\footnotesize $1.3\cdot10^1$}\\
{\footnotesize 158.68 } &  {\footnotesize $19200\times 8389$} & {\footnotesize $5.0\cdot10^6$} & {\footnotesize $1.7\cdot10^{10}$} & {\footnotesize $3.0\cdot10^{-4}$} & {\footnotesize $2.5\cdot10^0$}\\
{\footnotesize 264.46 } &  {\footnotesize $32000\times 5033$} & {\footnotesize $5.0\cdot10^6$} & {\footnotesize $8.1\cdot10^{10}$} & {\footnotesize $6.2\cdot10^{-5}$} & {\footnotesize $3.1\cdot10^{-1}$}\\
{\footnotesize 370.25 } &  {\footnotesize $44800\times 3595$} & {\footnotesize $5.0\cdot10^6$} & {\footnotesize $2.3\cdot10^{11}$} & {\footnotesize $2.2\cdot10^{-5}$} & {\footnotesize $7.9\cdot10^{-2}$}\\
{\footnotesize 528.93 } &  {\footnotesize $64000\times 2517$} & {\footnotesize $5.0\cdot10^6$} & {\footnotesize $6.9\cdot10^{11}$} & {\footnotesize $7.3\cdot10^{-6}$} & {\footnotesize $1.8\cdot10^{-2}$}\\
{\footnotesize 793.39 } &  {\footnotesize $96000\times 1678$} & {\footnotesize $5.0\cdot10^6$} & {\footnotesize $2.4\cdot10^{12}$} & {\footnotesize $2.1\cdot10^{-6}$} & {\footnotesize $3.5\cdot10^{-3}$}\\
{\footnotesize 1057.85 } &  {\footnotesize $128000\times 1258$} & {\footnotesize $5.0\cdot10^6$} & {\footnotesize $5.8\cdot10^{12}$} & {\footnotesize $8.6\cdot10^{-7}$} & {\footnotesize $1.1\cdot10^{-3}$}\\
{\footnotesize 1586.78 } &  {\footnotesize $192000\times 839$} & {\footnotesize $5.0\cdot10^6$} & {\footnotesize $2.0\cdot10^{13}$} & {\footnotesize $2.5\cdot10^{-7}$} & {\footnotesize $2.1\cdot10^{-4}$}\\
{\footnotesize 2115.70 } &  {\footnotesize $256000\times 629$} & {\footnotesize $5.0\cdot10^6$} & {\footnotesize $4.9\cdot10^{13}$} & {\footnotesize $1.0\cdot10^{-7}$} & {\footnotesize $6.3\cdot10^{-5}$}\\
\hline
\end{tabular}
\caption{
\label{tab:RaoulMCSimuls}
Details of the ring systems obtained by hierarchical crumpling (Sec.~\ref{sec:Fractals}).
$Z_r$: number of entanglements per single ring;
$N$: number of lattice bonds per single ring;
$M$: number of rings per each system;
$\tau_{tot}$: total length of MC simulation, expressed in single MC steps;
$\tau_{tot} / \tau_{eq}$: total number of independent MC configurations,
where $\tau_{eq}$ is the equilibration time estimated {\it via} the mean-square displacement of the rings center of mass. 
The first block of the table corresponds to direct construction of the ring systems and successive equilibration by extensive Monte Carlo (MC) simulations.
The second and third blocks correspond, respectively, to the first and second generation by employing hierarchical crumpling on rings from the first block 
with $Z_r\geq1.24$.
}
\end{table}

{\it Step 2} --
This step consists in a local Monte Carlo equilibration of the rings on the entanglement scale over a time scale of the order of $\tau_e$, where monomers diffuse over a distance of the order of $d_T = l_K \sqrt{N_{eK}/6}$ (Sec.~\ref{sec:Entanglement units}).
Compared to brute-force equilibration, this is a much faster procedure for long ring polymers, since the number of operations scales as $N \tau_e$,
which compares favorably to complete equilibration
(estimated to scale with ring size as $\sim N^{3.5}$). 
Computation time for the hierarchical model is insignificant (couple of days on a single GPU), even for the largest system ($256,000$ monomers, or $\approx 2100$ entanglement lengths).
Fig.~\ref{fig:hierarchical} (bottom row) illustrates the whole procedure of hierarchical crumpling starting from a ring with $Z_r\approx16$ (leftmost panel) to the final conformation with $Z_r\approx1000$ (rightmost panel).
A summary of the studied systems is given in Table~\ref{tab:RaoulMCSimuls}.

\section{Results}\label{sec:Results}

In our discrete models, individual ring conformations are characterised by the monomer positions, $\vec R_i$ with $i\in [1,N]$, or the bond vectors, $\vec b_i\equiv \vec R_{i+1} - \vec R_i$.
To account for periodic boundary conditions on the ring connectivity, it is convention to extend the range of admissible monomer indices to $\mathbb{Z}$ and define $\vec R_0 \equiv \vec R_N$, $\vec R_{N+1}\equiv \vec R_1$ and generally $\vec R_{i} \equiv \vec R_{\mod(i,N)}$ and $\vec b_{i} \equiv \vec b_{\mod(i,N)}$.
The spatial distance, $\vec R_{ij}\equiv \vec R_j - \vec R_i$, between any two points can be written as a sum over the interjacent bond vectors.
On a ring, ``interjacent'' can equally well be defined clockwise or counter-clockwise: $\vec R_{ij} = \sum_{k=i}^{j-1} \vec b_k = \sum_{k=j}^{i-1}\vec b_k$. 

To simplify comparisons between the different polymer models, we quote contour distances in units of length ($L$) and not in units of the corresponding number of bonds ($n$), $L=n l_b$, where $l_b=0.97\sigma$ for off-lattice fiber models and $l_b=0.714b$ for the hierarchical model. In particular, all results are expressed in entanglement units (Sec.~\ref{sec:Entanglement units}).

\begin{figure}
\includegraphics[width=0.50\textwidth]{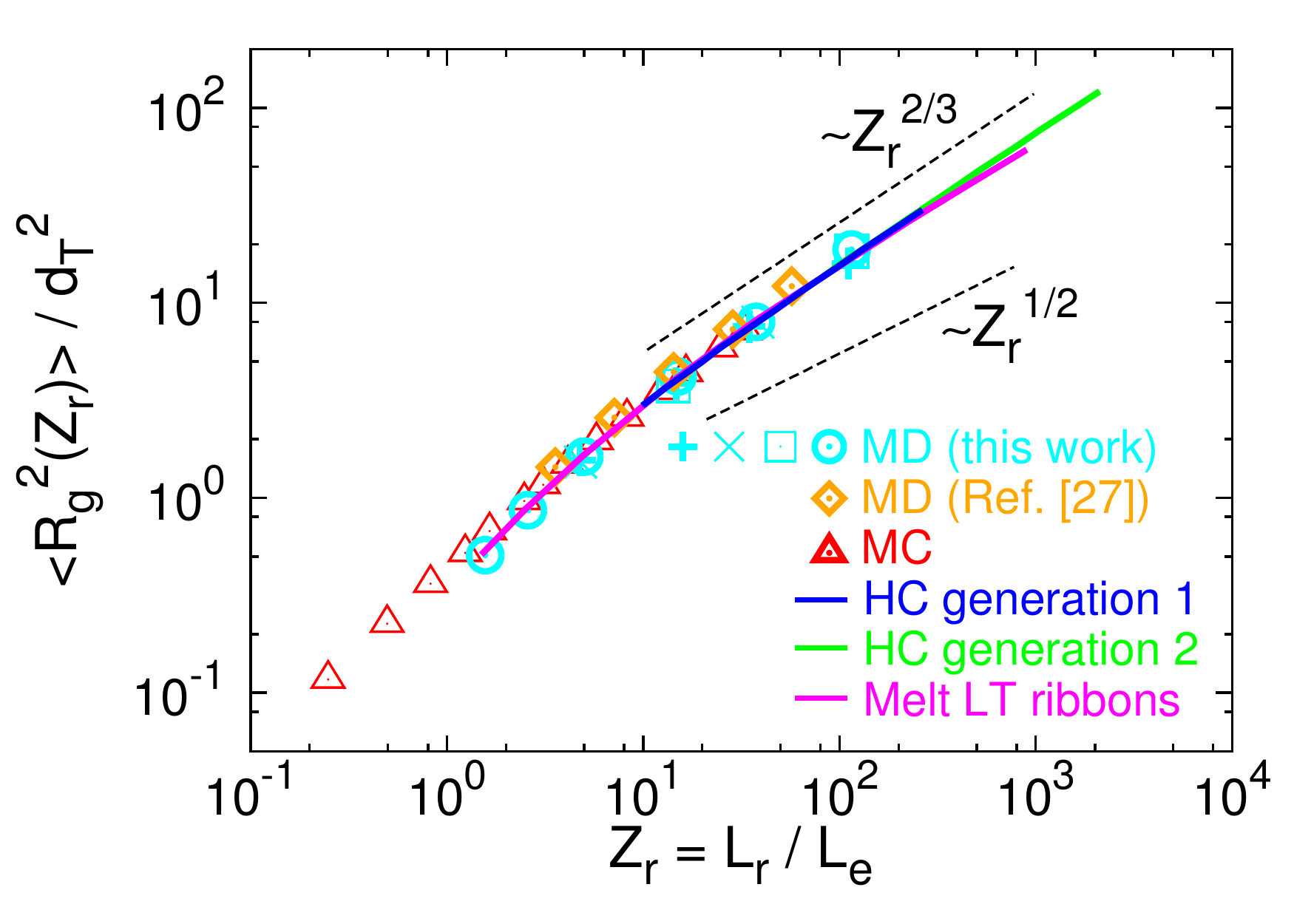}
\caption{
\label{fig:Rg2-ScalingPlot}
Ring mean-square gyration radius, $\langle R_g^2 \rangle$, as a function of ring contour length, $Z_r$. 
Lines with different colors correspond to the theoretically expected radii for the different types of structures generated, respectively, by hierarchical crumpling (HC, Sec.~\ref{sec:Fractals}) and as branched ribbons from interacting lattice trees in melt (Sec.~\ref{sec:Lattice model of double folded rings}).
MC symbols ($\bigtriangleup$) are for simulation data of melt structures equilibrated by Monte Carlo simulations of the elastic lattice polymer model (Sec.~\ref{sec:Elastic lattice polymer model}).
MD symbols are for simulation data of melt structures equilibrated by brute-force Molecular Dynamics simulations (Sec.~\ref{sec:fiberModel}) starting from the following initial states:
($+$) Klein-ribbons;
($\times$) Moore rings;
($\square$) Hilbert ribbons;
($\circ$) branched ribbons from ideal lattice trees;
($\Diamond$) data from Halverson {\it et al.}, Ref.~\cite{Halverson2011_1}.
}
\end{figure}

\subsection{Gyration radii}\label{sec:Rg}

The simplest overall measure of the spatial size of crumpled rings is their mean-square gyration radius:
\begin{equation}\label{eq:Rg}
\langle R_g^2(N) \rangle \equiv  \frac{1}{N} \sum_{i=1}^N \langle (\vec R_i - \vec R_{cm})^2 \rangle = \frac{1}{2N^2} \sum_{i,j=1}^N \langle \vec R_{ij}^2 \rangle \, . 
\end{equation}
where $\vec R_{cm} = \frac{1}{N}\sum_{i=1}^N \vec R_i$ denotes the ring centre of mass.

Fig.~\ref{fig:Rg2-ScalingPlot} shows a comparison of ``gold standard'' MD and MC data from Refs.~\cite{RosaEveraersPRL2014,Halverson2011_1,SchramBarkema2018} to results obtained for ring melts derived from lattice tree melts and hierarchical crumpling.
The excellent agreement between the different data sets validates the entanglement units (Sec.~\ref{sec:Entanglement units}) we employ to compare results for different models and provides a first set of evidence supporting the two construction algorithms. 
The crossover between the local to the crumpled regime is located around $Z_r\approx10$ entanglement lengths. 

\begin{figure}
\includegraphics[width=0.5\textwidth]{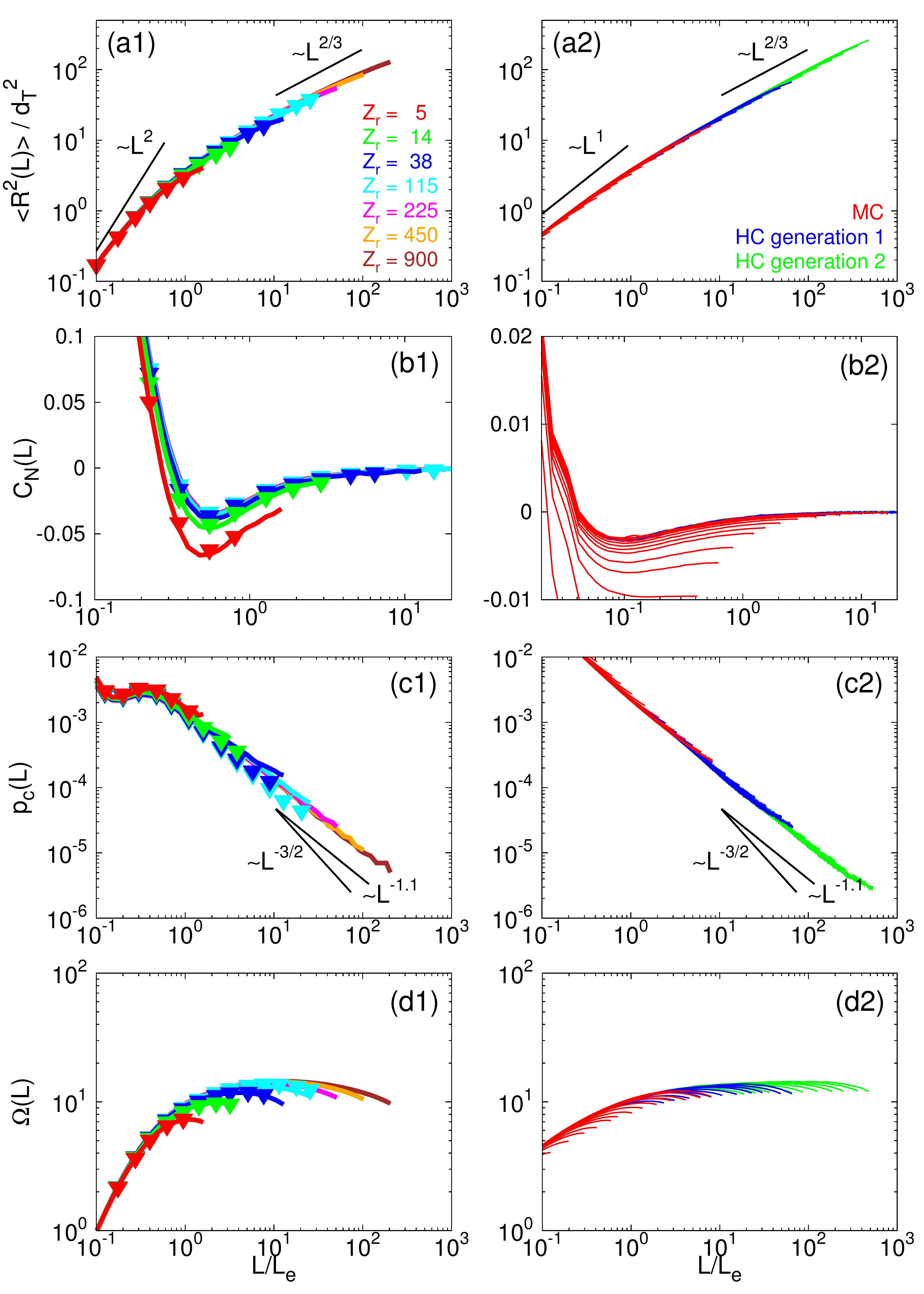}
\caption{
\label{fig:internal ring statistics}
More detailed measures of rings statistics.
L.h. column: Comparison between ``gold standard'' MD-equilibrated rings (symbols) and rings derived from interacting lattice tree melts (solid lines, same data as in Ref.~\cite{RosaEveraersPRL2014}).
R.h. column: Comparison between data for fully MC-equilibrated rings (red lines) to rings from melts generated via hierarchical crumpling (blue lines: first generation; green lines: second generation). In view of their higher statistical accuracy, we have in this case preferred to show them as lines. 
Panels (a1) and (a2): Mean-square internal distances, Eq.~(\ref{eq:R2L})
Panels (b1) and (b2): Bond-orientation correlation function, Eq.~(\ref{eq:bacf}).
Panels (c1) and (c2): Contact probabilities.
Panels (d1) and (d2): Overlap parameter, Eq.~(\ref{eq:OverlapParameter}).
}
\end{figure}
%

\subsection{Mean-square internal distances}\label{sec:R2}

More detailed information is provided by the mean-square internal distances between monomers as a function of their contour distance, $n= |i-j|$:
\begin{equation}\label{eq:R2L}
\rsqn = \frac{1}{N} \sum_{i=1}^N \langle \vec R_{i,i+n}^2 \rangle \, .
\end{equation}
For ring polymers with total contour length $N$, $\rsqn=\langle R^2(N-n)\rangle$. 
In panels (a1) and (a2) of Fig.~\ref{fig:internal ring statistics} we show corresponding data up to a contour distance of $n=N/4$, where ring closure plays only a small role. Again, there is excellent agreement between ``gold standard'' reference data and results obtained by the two construction methods. Again we have used entanglement units to facilitate the comparison between data for the two types of polymer models. 
The most notable difference occurs at small scales, where the lattice model exhibits random walk statistics, $\rsqn\sim n$, while the fiber model crosses over to rigid rod behavior, $\rsqn\sim n^2$.

\subsection{Correlation of the bond-vector orientation along the ring}\label{sec:BondBondCorrFunct}

The bond-vector orientation correlation function is defined by:
\begin{equation}\label{eq:bacf}
C_N(n) = \frac{1}{N} \sum_{i=1}^N \langle {\hat t}_i \cdot {\hat t}_{i+n} \rangle \, ,
\end{equation}
where ${\hat t}_i = \vec b_i / \vec | b_i |$ is the normalized bond vector. 
Note that both data sets shown in panels (b1) and (b2) of Fig.~\ref{fig:internal ring statistics} exhibit on the entanglement scale an anti-correlation, characteristic of looping,
which appears more pronounced for rings from melts of interacting lattice trees (notice the different ranges of the $y$-axes).

\subsection{Contact probabilities}\label{sec:pc}

The contact probabilities, $p_c(n)$, between monomers as a function of contour distance are shown in panels (c1) and (c2) of Fig.~\ref{fig:internal ring statistics}. Again, there is excellent agreement between the MC reference data and results for trees from melts generated via hierarchical crumpling.
In contrast, there are small but systematic deviations between the MD reference data for the fiber model and results for the lattice tree melt derived ensemble.
The employed contact radii were set to two times the bead diameter $\sigma$ for the fiber model and to the lattice constant for the lattice model. 
Both data sets are compatible with the power-law decay $p_c(L) \sim L^{-1.11 \pm 0.01}$ reported by Halverson {\it et al.}~\cite{Halverson2011_1}.

\subsection{Overlap parameter}\label{sec:Overlap}
The simplest measure of how different rings pack together, the so-called overlap parameter:
\begin{equation}\label{eq:OverlapParameter}
\Omega(L) \equiv \frac{\rho_K}{L/l_K} \langle R^2(L) \rangle^{3/2} \, ,
\end{equation}
expresses how many ring strands of linear size $L$ share the average volume spanned by each of them.
Notice, that both systems (panels (d1) and (d2) of Fig.~\ref{fig:internal ring statistics}) converge to the {\it universal}~\cite{KavassalisNoolandiPRL1987,FettersMacromolecules1994,uchida} entanglement threshold $\Omega(L \rightarrow \infty) \approx 20$.
For a reformulation of the argument in terms of a threshold for the reduced monomer self-density, $\hat\rho_{self}=0.5$, at the centers of mass of crumpled rings we refer the reader to Ref.~\cite{RosaEveraersPRL2014}. While the two definition essentially only differ by geometrical prefactors, the latter formulation provides an intuitive explanation for the binary character of entanglements~\cite{EveraersPRE2012}.

\begin{figure}
\includegraphics[width=0.5\textwidth]{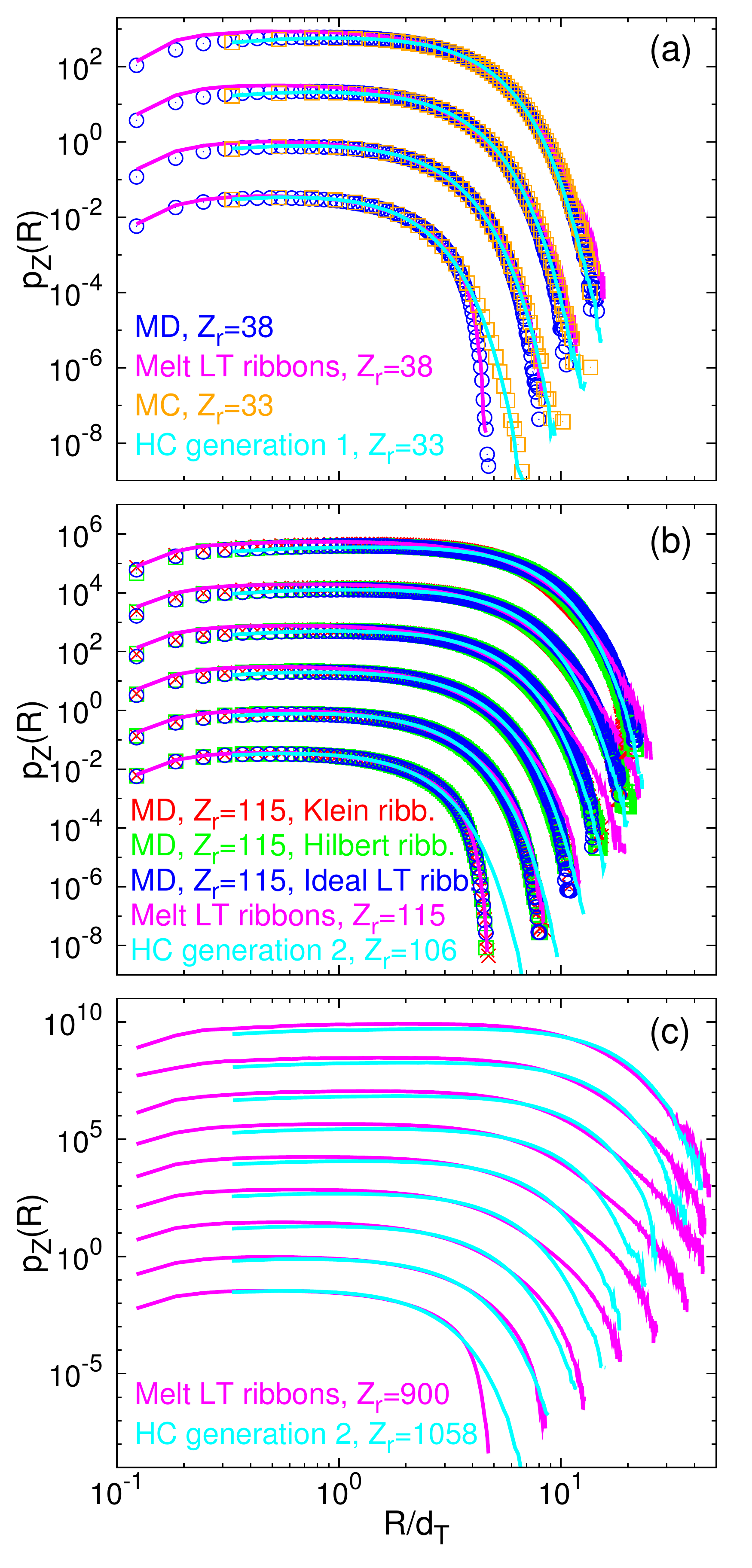}
\caption{
\label{fig:EndToEndPDFs}
Distribution functions of end-to-end internal distances, $p_Z(R)$, for chain contour lengths $Z$ in rings of total contour length $Z_r$.
Symbols and lines are, respectively, for ring structures obtained by brute-force equilibration (either MD or MC) and constructed according to the hierarchical crumpling (HC) method or as branched ribbons from melts of lattice trees (see legends).
In each panel, the different curves (from bottom to top) correspond to different contour lengths $Z$ and have been shifted by an arbitrary prefactor for better visualization:
(a)
Results for $Z=1,2,4,8$.
(b)
Results for $Z=1,2,4,8,16,32$.
(c)
Results for $Z=1,2,4,8,16,32,64,128,256$.
}
\end{figure}
%

\subsection{End-to-end distance distributions for ring sections}\label{sec:End2End}

The most detailed information on the single ring structure can be obtained from the distribution functions, $p_Z(R)$, of end-to-end distances $R$ of ring sections of contour length $Z=L/L_e$ in rings with total contour length $Z_r$.
Our results for $Z_r\approx 35,110,1000$ and $Z\le Z_r/4$ are summarized in the three panels of Fig.~\ref{fig:EndToEndPDFs}.

For $Z_r\approx 35$ (panel (a)) we have reliably equilibrated MD or MC ``gold standard'' data for both polymer models.
They are in perfect agreement except on the smallest investigated scale, $Z=1$, 
where the fluctuations in contour length allow the lattice model to reach larger extensions than the fiber model. 
Furthermore, there is perfect agreement between 
(i) data for rings derived from lattice tree melts and the corresponding MD reference data as well as
(ii) data for rings from melts generated via hierarchical crumpling and the corresponding MC reference data.

For $Z_r\approx 110$ (panel (b)) we have reasonably equilibrated MD ``gold standard'' data for the fiber model. Apart from the inevitable deviations for small $Z$ and large distances, there is again excellent agreement with the HC ensemble.
However, the lattice tree melts ensemble seems to contain a slightly higher proportion of strongly extended ring sections.

For $Z_r\approx 1000$ (panel (c)) we can only compare the HC and lattice tree melts ensembles.
Qualitatively, the observations confirm the trends observed for the smaller ring sizes. The overall agreement is good, but the extensions of the ring sections in the lattice tree melts ensemble exhibit a broader tail.

\section{Discussion}\label{sec:Discussion}

Our ``gold standard'' reference data for $Z_r \le {\cal O}(100)$ for the two different polymer models from Refs.~\cite{RosaEveraersPRL2014,SchramBarkema2018} are in good agreement with each other and previously reported data for untangled ring melts~\cite{mullerPRE1996,mullerPRE2000,Vettorel2009,DeguchiJCP2009,Halverson2011_1,HalversonPRL2012,Halverson2011_2}.
In particular, all results support the conjecture, that the rings crumple and asymptotically adopt compact conformations with $\nu=1/d$.
In the following, we focus on the comparison to the untangled melt ensembles, which we have derived from a wide range of theoretically inspired (lattice) models for the large scale structure of crumpled rings with $Z_r \le {\cal O}(1000)$.

\subsection{Hierarchical crumpling and lattice tree melts {\it ex aequo}?}
Our detailed comparison to the reference data appears to suggests that two construction algorithms pass the ``Feynman test'' nearly {\it ex aequo}:
(i) the assembly of untangled ring melts on the basis of lattice tree melts
and
(ii) hierarchical crumpling. 
Ring melts derived from Klein ribbons, Moore curves, Hilbert ribbons, and ideal randomly branching trees~\cite{RosaEveraersPRL2014} fail more or less clearly as illustrated by the detailed analysis of their conformational statistics, see Figs.~S\ref{fig:Klein}-S\ref{fig:IdealLatticeAnimal} in the SM.

The success of the lattice tree method provides insight into the mechanisms of entropy maximisation in untangled polymers.
In particular, it supports mechanical analogies which allow for further theoretical analysis.  
We found strong evidence for the scenario, that rings crumple by adopting ribbon structures characterized by randomly branched looping on the entanglement scale~\cite{KhokhlovNechaev85,RubinsteinPRL1986,ORD_PRL1994}.
While we observe ideal lattice tree behavior on scales up to $Z_r=30$ entanglements (compare l.h.s panels of Fig.~\ref{fig:internal ring statistics} to Fig.~S\ref{fig:IdealLatticeAnimal} in SM),
the approximation breaks down {\it before} the characteristic $\nu=1/4$ regime is reached (for a discussion on the crossover, see Ref.~\cite{RosaEveraersJCP2016}).
The observed swelling of the randomly branched loop structures of the rings is in very good agreement
with the behavior of {\it interacting} randomly branching chains (or lattice trees) in a melt~\cite{DaoudJoanny1981,KhokhlovNechaev85,GrosbergSoftMatter2014,EveraersRosaFloryReview2017,RosaEveraersJCP2016} and hence supports the conjecture,  that crumpled rings are asymptotically compact, $\nu=1/d$.

From an algorithmic point of view, ``hierarchical crumpling'' is probably best seen as a simple multi-grid~\cite{MultiGridMethods} method,
which speeds up the relaxation of the larger scales by simulating them with a much coarser discretisation than the actual polymer model. 
Our particular implementation for the elastic lattice polymer realises an ideal situation from the (Monte Carlo) Renormalization Group~\cite{NewmanBarkemaMC-Book} point of view: no matter the degree of coarse-graining, the system is on all scales described by exactly the same (lattice) model. 
The success of the ``hierarchical crumpling'' scheme provides insight into the fractal nature of the problem:
untangled rings crumple to adopt sizes close to the entanglement threshold~\cite{RosaEveraersPRL2014}, because the same physics of entropy maximisation under topological constraints governs their behavior on all scales beyond the entanglement scale~\cite{PanyukovRubinsteinMacromolecules2016}. 

The fractal picture of the entanglement constraints in crumpled rings and the mechanistic interpretation of crumpling through the tree analogy are linked through the prescription, that double folding and branching occur on the entanglement scale. 
However, the good agreement between the ``gold standard'' reference ensembles and the untangled melt ensembles derived from ``lattice tree melts'' and via ``hierarchical crumpling'' does {\em not} mean, that crumpling is {\em fully} described through double folding, branching and swelling.

\begin{figure}
\includegraphics[width=0.50\textwidth]{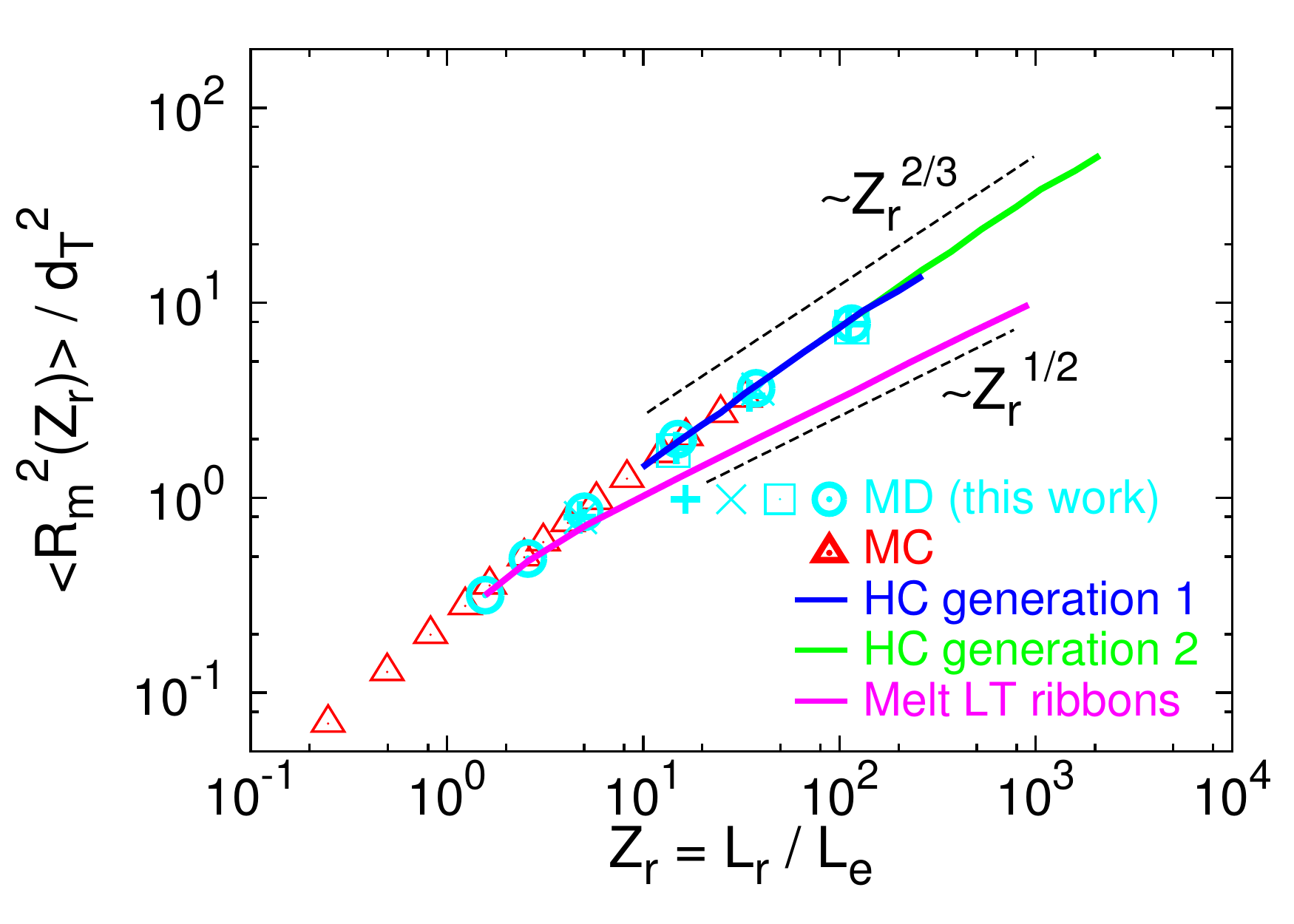}
\caption{
\label{fig:Rm2-ScalingPlot}
Mean-square magnetic radius ($\langle R_m^2 \rangle$) as a function of ring contour length, $Z_r$.
Lines, symbols and colors are as in Fig.~\ref{fig:Rg2-ScalingPlot}.
}
\end{figure}
%

\subsection{Loop opening and the magnetic radius}
Visual inspection of the randomly chosen rings in Fig.~\ref{fig:hierarchical} immediately points to a major difference between the two ensembles:
the presence of a large open loop in a ring conformation generated by hierarchical crumpling is not compatible with a fully double-folded structure.
In the following, we investigate a new observable inspired by an analogy to magnetostatics, which provides a new and convenient measure of the statistical significance of partial ring opening.

Consider the dipole moment, $\vec m$, characterising the magnetic far field generated by a loop carrying a constant electric current, $I$.
Following the work of Amp\`ere~\cite{JacksonElectrodynamics},  $\vec m = I \vec A$ where
\begin{equation}\label{eq:EnclosedArea}
\vec A = \frac12 \sum_{i=1}^N \vec r_i \times (\vec r_{i+1}-\vec r_i) = \frac12 \sum_{i=1}^N \vec r_i \times \vec r_{i+1} 
\end{equation}
for a piecewise straight line like our ring polymers. 
Evaluation for the simple geometry of a planar circle shows that the magnitude of $\vec A$ indicates the enclosed area, suggesting to define a ``magnetic'' ring radius as
\begin{equation}\label{eq:MagnRadius}
R_m^2 = \frac1\pi \left| \vec A \right| \ .
\end{equation}

We have evaluated $A$ and $R_m$ for all our data sets.
Our results for the magnetic radius are shown in Fig.~\ref{fig:Rm2-ScalingPlot}. To ease the comparison, we have used the same representation as for the gyration radii in Fig.~\ref{fig:Rg2-ScalingPlot}. The temporal evolution of $\langle R_m^2 \rangle$ and $\langle R_g^2 \rangle$ for the various starting states in our simulations of the fiber model is further illustrated in Fig.~S\ref{fig:Rg2andRm2Equilibration} in the SM. 
In particular, we find for crumpled rings $R_m^2 \propto R_g^2 \propto Z_r^{2/3}$. 

The proportionality of these different measures of the ring size perfectly illustrates the fractal character of the ring structures emerging in untangled melts.
This feature is also reproduced by Moore curves, which otherwise fail to reproduce the statistics of crumpled rings in any quantitative sense (Fig.~S\ref{fig:Moore} in the SM),
even though they were constructed to faithfully represent the molecular volume accessible the rings.
Remarkably, the gyration {\em and} magnetic radii both perfectly coincide with the reference data for rings from melts generated via hierarchical crumpling. 
In contrast, for rings derived from lattice tree melts 
$R_m^2  \propto Z_r^{1/2}$, with enclosed areas which are substantially smaller than for the reference data.

\begin{figure}
\includegraphics[width=0.50\textwidth]{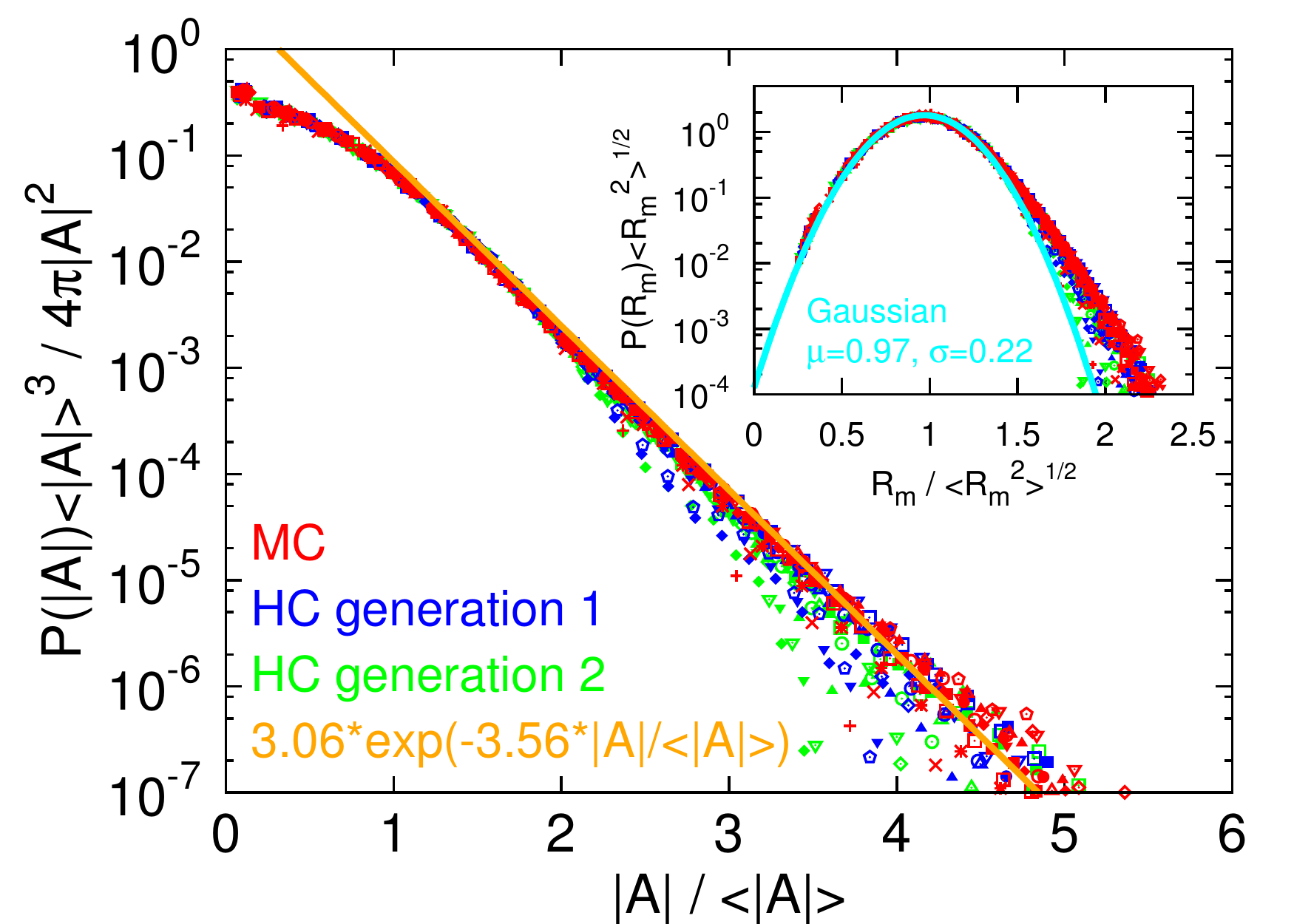}
\caption{
\label{fig:Amagsq_dist}
Probability distribution function, $q\left(\frac{\vec A}{\langle |\vec A| \rangle}\right)$, of the vector representing the total area, $\vec A$ (Eq.~(\ref{eq:EnclosedArea})), enclosed by crumpled ring polymers.
Symbols are for MC data (red) and data from the first (blue) and the second (green) generation of hierarchical crumpling (HC) .
The solid line is the result of fitting the long-tail $(2.5\leq |\vec A| / \langle|\vec A|\rangle \leq5.0)$ behavior to an exponential function.
Inset: Probability distribution function, $P(R_m)$, of the magnetic radius, $R_m$ (Eq.~(\ref{eq:MagnRadius})).
The solid line is the result of fitting the data around the maximum $(0.5\leq R_m / \langle R_m^2 \rangle^{1/2} \leq1.5)$ to the Gaussian function with mean $\mu$ and standard deviation $\sigma$.
}
\end{figure}

For the further analysis we focus on the elastic lattice polymer model, where we have data of much better statistical quality. 
As a first step, we analyze the distribution functions of $\vec A$ and $R_m$. 
Our starting point is the sampled distribution, $P(| \vec A|)$, of the magnitude of the enclosed surface. 
From this, we can derive the distribution $p(\vec A)$ of the underlying vector $\vec A$, 
\begin{equation}
p(\vec A) = \frac{P(|\vec A|)}{4\pi |\vec A|^2} \ .
\label{eq:pr_of_A}
\end{equation}
To compare results obtained for different ring sizes, it is useful to consider corresponding distribution of rescaled surfaces:
\begin{eqnarray}
q\left(\frac{\vec A}{\langle |\vec A| \rangle}\right) = \frac{\langle |\vec A| \rangle^3} {4\pi |\vec A|^2} P(|\vec A|) \, .
\label{eq:q_of_A}
\end{eqnarray}
Fig.~\ref{fig:Amagsq_dist} shows the expected collapse of data for $q\left(\frac{\vec A}{\langle |\vec A| \rangle}\right)$ for different generations of the hierarchical crumpling scheme.
Furthermore, our results suggests that the distributions decay exponentially (solid line) for large $ |\vec A| $.
In particular, $\sqrt{\langle  \vec A\cdot\vec A \rangle} / \langle |\vec A| \rangle \approx 1.1$. 

The corresponding distributions for the magnetic radius (see Fig.~\ref{fig:Amagsq_dist}, inset) are given by:
\begin{eqnarray}
P\left(R_m=\sqrt{|\vec A|/\pi} \right) &=& \sqrt{ 4\pi |\vec A|} \ P(|\vec A|) \label{eq:P_of_Rm}\\
Q \left( \frac{R_m}{\langle R_m^2 \rangle^{1/2}} = \sqrt{\frac{|\vec A|}{\langle |\vec A| \rangle}} \right) 
    &=&  \sqrt{ 4 |\vec A|\, \langle |\vec A| \rangle} \ P(|\vec A|) \ . \label{eq:Q_of_Rm}
\end{eqnarray}
They are nearly Gaussian with a peak close to the root-mean-square magnetic radius (solid line).

\subsection{The emerging picture}
It is easy to understand, why the tree melt derived structures fail the ``Feynman test'' so badly with respect to the magnetic radius, even though they perfectly reproduce other structural observables like the gyration radius. By construction, $A \equiv 0$ for our starting states of tightly double-folded rings. The enclosed surface is zero and each primitive path segment is occupied by  two oppositely oriented ring segments, whose ``currents'' therefore exactly cancel.
Local equilibration over $\tau_e$ opens $Z_r/2$ randomly oriented surface elements $\vec A_i$ of the order of $d_T^2$.
For the expected minimal surface,
\begin{eqnarray}
\langle A_{min} \rangle 
  \sim \sum_{i=1}^{Z_r/2} |\vec A_i| 
   \sim d_T^2 Z_r,
\end{eqnarray}
double folded rings should still pass the ``Feynman test'', at least on a scaling level~\cite{SmrekGrosbergACSMacroLett2016}.
However, they fail for the magnetic radius, which derives from a {\em vector} sum over the differently oriented surface elements,
\begin{eqnarray}
\langle R_m^2 \rangle 
   \sim \sqrt{\langle  \vec A\cdot\vec A \rangle}
  \sim \sqrt{\sum_{i=1}^{Z_r/2} |\vec A_i|^2} 
   \sim d_T^2 Z_r^{1/2}\ .
\end{eqnarray}
In particular, the above result holds {\em independently} of the branching and conformational statistics of the primitive chain or tree.
This is well borne out by our data for Klein ribbons, ideal and interacting lattice trees in the panel for $t=\tau_e$ in Fig.~S\ref{fig:Rg2andRm2Equilibration} in the SM. 

But what does all this say about untangled polymer melts?
How can we understand 
(i) that most aspects of the statistics (and dynamics) of crumpled rings appear to be in excellent agreement with the idea of double-folding and the analogy to lattice tree melts,
(ii) that there is ample anecdotic proof of the opening of larger loops through visual inspection of crumpled ring conformations ({\it e.g.}, Fig.~\ref{fig:hierarchical}), and
(iii) that the magnetic radius appears to scale like the ring gyration radius, while the enclosed minimal surface grows linearly with the rings size~\cite{SmrekGrosbergACSMacroLett2016}? 

A naive interpretation is to assume that crumpled rings open  ${\cal O}(1)$ loops of a spatial size of the same order as their gyration radius.
The rings being territorial, loops of this size can avoid topological linkage, since in this scenario their overall concentration corresponds to the overlap concentration~\cite{PanyukovRubinsteinMacromolecules2016}.
If we assume Gaussian statistics for open loops, their creation requires with $n \propto N^{2/3}$ monomers only a fraction, $n/N \propto N^{-1/3}$, of the total ring mass.
This would explain, why open loops make no significant contribution to a wide range of static observables, where the reference data is in excellent agreement with results for the lattice tree melt derived ensemble (Figs.~\ref{fig:Rg2-ScalingPlot} to~\ref{fig:EndToEndPDFs} and Ref.~\cite{RosaEveraersPRL2014}).
Furthermore, and in agreement with our present findings and those from Ref.~\cite{SmrekGrosbergACSMacroLett2016},
the presence of such relatively large open loops would dominate the magnetic radius,
\begin{eqnarray}
\langle R_m^2 \rangle 
  \sim \sqrt{d_T^4 Z_r + \left( d_T^2 Z_r^{2/3}\right)^2} 
   \sim d_T^2 Z_r^{2/3}\ ,
\end{eqnarray}
but hardly affect the enclosed minimal surface:
\begin{eqnarray}
\langle A_{min} \rangle 
   \sim d_T^2 Z_r \left(1 - Z_r^{-1/3} \right) + d_T^2 Z_r^{2/3} \approx d_T^2 Z_r\ .
\end{eqnarray}
Given the fractal ring structure, corresponding openings can also exist on smaller scales all the way down to the entanglement scale. In this limit, the difference between open and double-folded sections vanishes, while the loop fraction formally approaches one. 
Refs.~\cite{ObukhovWittmerEPL2014,PanyukovRubinsteinMacromolecules2016} discuss corresponding scenarios including two different proposals for distribution functions for the loop sizes.
It would be interesting to see, if these approaches can be generalised to predict the orientational correlations between the loops required to estimate the overall magnetic radius of crumpled rings.

\section{Summary and Conclusion}\label{sec:Concls}
We have extended the multi-scale approach of Ref.~\cite{RosaEveraersPRL2014} for studying dense solutions of untangled ring polymers.
The employed ``gold standard'' reference data for $Z_r \le {\cal O}(100)$ were obtained by brute-force equilibration of an off-lattice fiber model~\cite{RosaEveraersPRL2014} and of a highly efficient elastic lattice polymer model~\cite{SchramBarkema2018}. 
They are in good agreement with the previously reported~\cite{mullerPRE1996,mullerPRE2000,Vettorel2009,DeguchiJCP2009,Halverson2011_1,HalversonPRL2012,Halverson2011_2} results for the statics and dynamics of untangled ring melts and support the conjecture, that crumpled rings are asymptotically compact, $\nu=1/d$.
In addition to the commonly analysed standard measures of (ring) polymer statistics, we have used an analogy to electrodynamics to define an easily calculable ``magnetic radius'', $R_m$, representing the area enclosed by a ring polymer (Eqs.~(\ref{eq:EnclosedArea}) and (\ref{eq:MagnRadius})). 
While $R_m\equiv 0$ for (tightly) double-folded rings, we found $R_m \propto R_g$ and hence $R_m \propto Z_r^{1/d}$ for well equilibrated crumpled rings.

The focus of our study lay on the detailed comparison of our reference structures to conformations derived from theoretically inspired (lattice) models for untangled rings. 
The considered structures range from fractal space filling curves (Fig.~S\ref{fig:MooreGenerations} in SM) over lattice tree melts (Fig.~\ref{fig:hierarchical}, top row) to ring melts constructed by a process that we have dubbed ``hierarchical crumpling'' (Fig.~\ref{fig:hierarchical}, bottom row).
The construction algorithms being computationally much more efficient than the brute-force equilibration of the original polymer models, we were able
(i) to generate model-derived ensembles of untangled melts for ring sizes $Z_r \le {\cal O}(1000)$
and
(ii) to implement a detailed ``Feynman test'' of our ability to construct (nearly) equilibrated conformations.
This allowed us to pursue two objectives: the identification of the physics underlying the crumpling of rings and the validation of multi-scale algorithms for generating plausible untangled melt structures for otherwise inaccessible ring sizes.

The success of the lattice tree melt analogy provides evidence (Figs.~\ref{fig:Rg2-ScalingPlot} to~\ref{fig:EndToEndPDFs}) for a mechanistic description of crumpling in terms of randomly branched, double-folded ring structures~\cite{KhokhlovNechaev85,RubinsteinPRL1986,ORD_PRL1994,GrosbergSoftMatter2014,SmrekGrosbergJPCM2015}.
The success of the hierarchical crumpling scheme illustrates that crumpling is governed on all scales by the same physics of entropy maximisation under topological constraints, which keeps the ring extensions close to the entanglement threshold~\cite{RosaEveraersPRL2014,PanyukovRubinsteinMacromolecules2016}. 
In theoretical physics, lattice animals and trees are often used interchangeably, because they fall into the same universality class and are hence characterized by the same critical exponents~\cite{ParisiSourlasPRL1981}.
Our results show how well a similar analogy works in the present context.
Nevertheless, there are differences as revealed by our analysis of the opening of larger loops and the analogous results in Ref.~\cite{SmrekKremerRosaACSML2019}.
For all its utility in {\em understanding} untangled melts, the lattice tree melt analogy thus fails the ``Feynman test'' of providing a recipe for {\it constructing} equilibrated untangled ring melts. 
Hierarchical crumpling achieves this goal remarkably well, at least within the range of observables we have investigated so far.

{\it Acknowledgements} --
This work was in part funded by the French National Research Agency (ANR-15-CE12-0006 -- ``EpiDevoMath'').
AR and RE acknowledge long and stimulating discussions with M. Rubinstein, A. Grosberg and M. Kolb. 
Simulations were performed  employing the computer facilities of the FLMSN, notably of the P\^ole Scientifique de Mod\'elisation Num\'erique (PSMN) and the Centre Blaise Pascal (CBP) at the Ecole Normale Sup\'erieure de Lyon. 
%


\begin{thebibliography}{78}
\bibitem[{Moo()}]{Moore_note}
\bibinfo{note}{This is not an artifact of the particular form of the Moore
  curve, but a consequence of the more general construction principle of
  filling space by subsequently occupying neighboring cubic blocks. Let $a$
  denote the linear dimension of such a block, which contains a chain contour
  length of $L(a)$. By construction, there is only a single chain of length
  $L(a)$ contained in the volume $a^3$, so that the contour length density is
  $\rho_{cl}=L(a)/a^3$. If the path enters the block at one corner and exits
  through a neighboring corner at a distance $\rsq{L(a)}=a^2$, across a face
  diagonal at a distance $\rsq{L(a)}=2a^2$, or across a body diagonal at a
  distance $\rsq{L(a)}=3a^2$, then the overlap parameter on this length scale
  is given by $1^{3/2}=1$, $2^{3/2}\approx2.8$, and $3^{3/2}\approx5.2$
  respectively.}
\bibitem[{Hil()}]{HilbertRibbons_note}
\bibinfo{note}{It is easy to show that the overlap parameters between Moore
  rings and Hilbert ribbons differ by a factor of $\approx 2$, and that similar
  relations should hold for other space-fillings curves which are constructed
  by subsequently occupying neighboring cubic blocks. Let $a$ denote the linear
  dimension of such a block, which contains a chain contour length of
  $L(a)$~\cite{Moore_note}. For analogs of the Hilbert ribbons the contour
  length density is $\rho_{cl} = 2 \, L(a)/a^3$, since the volume contains two
  independent chain sections. As a consequence, the overlap parameter increases
  by a factor of two.}
\end{thebibliography}


\clearpage

\setcounter{section}{0}
\setcounter{figure}{0}
\setcounter{table}{0}
\setcounter{equation}{0}

\renewcommand{\figurename}{Fig. S}
\renewcommand{\tablename}{Table S}

{\large \bf Supplemental Material}

\tableofcontents


\section{Model ring melt conformations derived from regular fractal space-filling curves}\label{sec:SM Fractals}

%
\begin{table}
\begin{tabular}{ccccc}
Initial state & $Z_r$ & $N \times M \times \#\mbox{RUNS}$ & $\tau_{tot} [\tau_{LJ}]$ & $\tau_{tot} / \tau_{eq}$\\
\hline
\hline
{\footnotesize Klein ribbon (I)} & {\footnotesize 4.8} &  {\footnotesize $190\times1\times1^*$} & {\footnotesize $1.2 \times 10^7$} & {\footnotesize $2 \times 10^3$}\\
{\footnotesize Klein ribbon (II)} & {\footnotesize 4.8} &  {\footnotesize $194\times1\times(\gtrsim100)^*$} & {\footnotesize $1.2 \times 10^7$} & {\footnotesize $2 \times 10^3$}\\
{\footnotesize Moore ring} & {\footnotesize 4.8} & {\footnotesize $192\times8\times1$} & {\footnotesize $1.2 \times 10^8$} & {\footnotesize $2 \times 10^4$}\\
{\footnotesize Ideal LT ribbon} & {\footnotesize 5.0} & {\footnotesize $200\times32\times1$} & {\footnotesize $1.2 \times 10^7$} & {\footnotesize $2 \times 10^3$}\\
\hline
{\footnotesize Klein ribbon (I)} & {\footnotesize 14.7} & {\footnotesize $589\times1\times1^*$} & {\footnotesize $1.2 \times 10^8$} & {\footnotesize $1.5 \times 10^3$}\\
{\footnotesize Klein ribbon (II)} & {\footnotesize 14.5} & {\footnotesize $582\times1\times(\gtrsim100)^*$} & {\footnotesize $1.2 \times 10^7$} & {\footnotesize $1.5 \times 10^2$}\\
{\footnotesize Hilbert ribbon} & {\footnotesize 14.3} & {\footnotesize $570\times8\times1$} & {\footnotesize $1.2 \times 10^8$} & {\footnotesize $1.5 \times 10^3$}\\
{\footnotesize Ideal LT ribbon} & {\footnotesize 15.0} & {\footnotesize $600\times8\times1$} & {\footnotesize $1.2 \times 10^8$} & {\footnotesize $1.5 \times 10^3$}\\
\hline
{\footnotesize Klein ribbon (I)} & {\footnotesize 34.7} &  {\footnotesize $1388\times1\times1^*$} & {\footnotesize $1.2 \times 10^8$} & {\footnotesize $1 \times 10^2$}\\
{\footnotesize Klein ribbon (II)} & {\footnotesize 34.9} &  {\footnotesize $1396\times1\times(\gtrsim100)^*$} & {\footnotesize $1.2 \times 10^7$} & {\footnotesize $1 \times 10^1$}\\
{\footnotesize Moore ring} & {\footnotesize 38.4} &  {\footnotesize $1536\times8\times1$} & {\footnotesize $2.4 \times 10^8$} & {\footnotesize $2 \times 10^2$}\\
{\footnotesize Ideal LT ribbon} & {\footnotesize 37.6} & {\footnotesize $1502\times16\times1$} & {\footnotesize $1.2 \times 10^8$} & {\footnotesize $1 \times 10^2$}\\
\hline
{\footnotesize Klein ribbon (I)} & {\footnotesize 110.8} & {\footnotesize $4433\times1\times1^*$} & {\footnotesize $1.2 \times 10^9$} & {\footnotesize ${\cal O}(10)$}\\
{\footnotesize Klein ribbon (II)} & {\footnotesize 110.2} & {\footnotesize $4409\times1\times(\gtrsim100)^*$} & {\footnotesize $1.2 \times 10^7$} & {\footnotesize ${\cal O}(0.1)$}\\
{\footnotesize Hilbert ribbon} & {\footnotesize 115.5} & {\footnotesize $4620\times8\times1$} & {\footnotesize $6.0 \times 10^8$} & {\footnotesize ${\cal O}(5) $}\\
{\footnotesize Ideal LT ribbon} & {\footnotesize 115.1} & {\footnotesize $4605\times8\times1$} & {\footnotesize $1.2 \times 10^8$} & {\footnotesize ${\cal O}(1) $}\\
\hline
{\footnotesize Moore ring} & {\footnotesize 307.2} & {\footnotesize $12288\times8\times1$} & {\footnotesize $1.2 \times 10^8$} & {\footnotesize --}\\
\hline
{\footnotesize Hilbert ribbon} & {\footnotesize 925.6} & {\footnotesize $37024\times1\times1$} & {\footnotesize $1.2 \times 10^8$} & {\footnotesize --}\\
\hline
\end{tabular}
\caption{
\label{tab:MDruns}
Details of the ring systems studied by Molecular Dynamics computer simulations (Sec.~\ref{sec:fiberModel} in the main text).
Initial states correspond to
(a)
the lattice models described in Sec.~\ref{sec:Lattice model of double folded rings} in the main text
and
(b)
the fractal models described in Sec.~\ref{sec:SM Fractals} here.
$Z_r$: number of entanglements per single ring;
$N$: number of Lennard-Jones monomers per single ring;
$M$: number of rings per each system;
$\#\mbox{RUNS}$: total number of independent MD runs;
$\tau_{tot}$: time-length of the single MD trajectory, expressed in elementary Lennard-Jones (LJ)~\cite{lammps,KremerGrestJCP1990} time steps ($\tau_{LJ}$);
$\tau_{tot} / \tau_{eq}$: total number of independent MD configurations, where
$\tau_{eq}$ is the {\it diffusion} equilibration time corresponding to chain motion beyond the polymer mean gyration radius. 
$^*$There are two sets of MD simulations for Klein ribbons.
Set I includes data from Ref.~\cite{RosaEveraersPRL2014} and is used for
Figs.~\ref{fig:Rg2-ScalingPlot}, \ref{fig:EndToEndPDFs} and~\ref{fig:Rm2-ScalingPlot} in the main text
and
Figs.~S\ref{fig:Klein} and~S\ref{fig:KleinHilbertRelaxation} here. 
Set II includes data consisting of, at least, $\#\mbox{RUNS}$ independent simulations of one-ring systems and is used for Fig.~S\ref{fig:Rg2andRm2Equilibration} here.
}
\end{table}
\begin{figure*}
\vspace{5mm}
\includegraphics[width=0.8\textwidth]{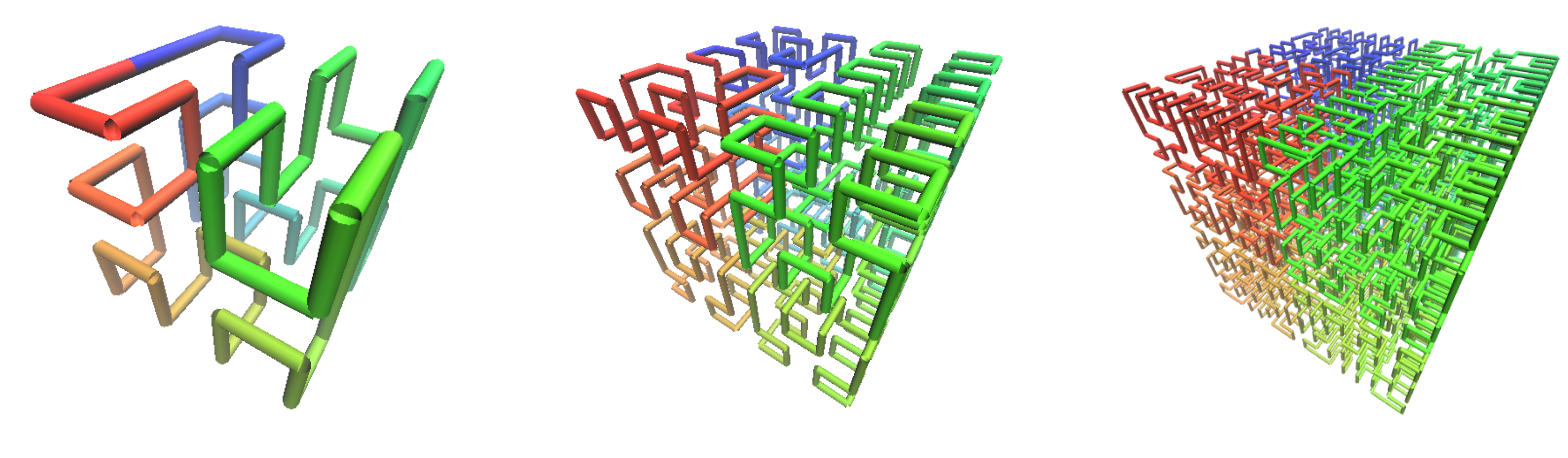}
\caption{
\label{fig:MooreGenerations}
Moore curves. From left to right are shown the first ($Z_r=4.8$), second ($Z_r=38.4$) and third ($Z_r=307.2$) generation of a single ring polymer.
}
\end{figure*}
{\bf Moore rings} --
The Moore ring is the closed version of the Hilbert curve and both can be obtained by a recursive numerical algorithm~\cite{SpaceFillingCurvesSagan}.
Bead-spring Moore rings are constructed by simply arranging the monomers along the contour line of the curve. 
With the additional constraint of monomer density $\rho = 0.1/\sigma^3$, admissible contour lengths $Z_r$ for Moore rings occupy a cubic box of volume, $V$, given by:
\begin{equation}\label{eq:MooreValues}
V = \frac{6^{3/2}}{20} \, d_T^3 \, Z_r \, .
\end{equation}
$Z_r$ is thus a multiple of 8 of $Z_0 \approx \frac{64 (\rho_K l_K)^{-1/2}}{L_e} \approx 5$, which leads to
$Z_r = L_r/L_e = 5, 38, 307$ (see Table~S\ref{tab:MDruns} and Fig.~S\ref{fig:MooreGenerations}).

{\bf Hilbert ribbons} --
As a hybrid between the fractal and the double-folded building strategies, we considered compact ribbon conformations where the ribbon axis follows a Hilbert curve instead of a random walk.
Hilbert ribbons 
are built according to a procedure analogous to the construction of Klein folded rings,
where now the contour length consists of a Hilbert curve.
Analogously to Moore rings, Hilbert ribbons occupy a volume $V$ given by Eq.~(\ref{eq:MooreValues}), where $Z_r$ is a multiple of 8 of
$Z_0 \approx \frac{128 (\rho_K l_K/2)^{-1/2}}{L_e} \approx 14$.
This leads to $Z_r = L_r/L_e = 14, 116, 926$ (see Table~S\ref{tab:MDruns}).

Moore rings and Hilbert ribbons have been also used as starting conformations for MD computer simulations.
The total number, $M$, of chains considered is summarized in Table~S\ref{tab:MDruns}.

\section{Properties of ring melts constructed from different polymer models: large scale structure}\label{sec:SupplMat:RingsStructure}
In the main text, we have focused on the ring melts derived from lattice tree melts or via hierarchical crumpling. To appreciate the success of these methods, it is useful to compare them to other plausible, but less refined models for crumpled rings, which we have studied in Ref.~\cite{RosaEveraersPRL2014}. Below, we provide more details and analyze the same observables
$(\langle R^2(L) \rangle, C_N(L), p_c(L), \Omega(L))$ as in the main text.

\begin{figure}
\includegraphics[width=0.48\textwidth]{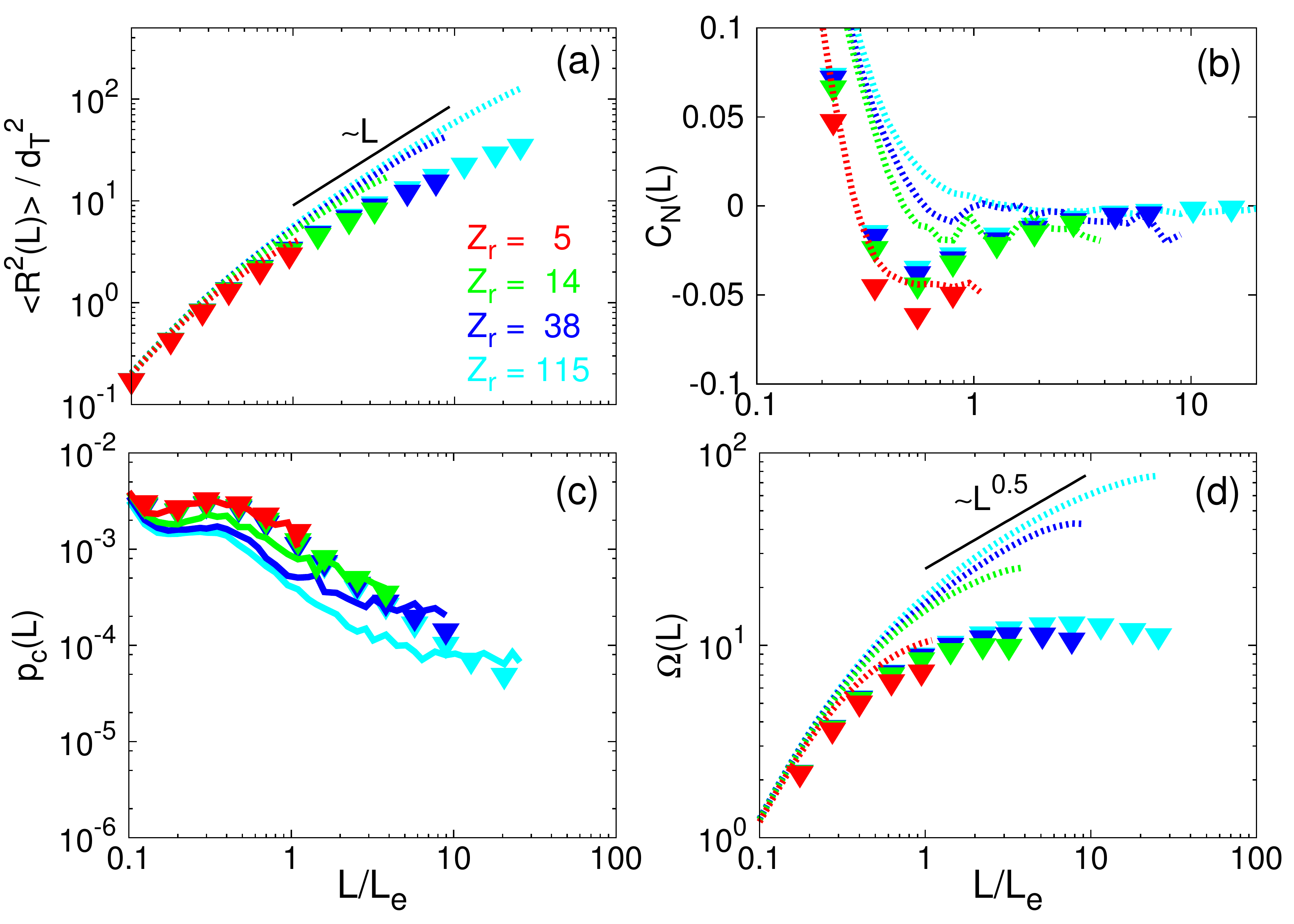}
\caption{
\label{fig:Klein}
Comparison of the conformational statistics for MD-equilibrated rings (symbols) and Klein ribbons
(lines:
dotted, average at $t=0$;
solid, average at $t=\tau_e$).
(a)
$\langle R^2(L) \rangle$,
Mean-square internal distances.
(b)
$C_N(L)$:
Bond-vector orientation correlation function.
(c)
$p_c(L)$:
Mean contact probability between monomers for contact distance $\leq 2\sigma$.
(d)
$\Omega(L)$:
Overlap parameter.
Data extend up to 1/4 of the corresponding rings contour lengths.
The same observables are used in Fig.~\ref{fig:internal ring statistics} in the main text and Figs.~S\ref{fig:Moore}-S\ref{fig:IdealLatticeAnimal}.
}
\end{figure}

{\bf Klein ribbons} -- 
For a ribbon axis with the same Kuhn length as in the fiber model, the conformational statistics of tightly wrapped rings turns out to be in almost perfect agreement with the corresponding Gaussian rings~\cite{RosaEveraersPRL2014}:
in particular, the mean-square internal distances obtained for the constructed Klein ribbons (dotted lines in Fig.~S\ref{fig:Klein}a)
are equivalent to the Gaussian ring law, $\rsqL = l_K L \left( 1-\frac{L}{L_r} \right)$, 
where the average is taken over all monomers of a ribbon with vanishing diameter and an axis with the same stiffness / Kuhn length as the chains.
As a consequence, there is also perfect agreement for quantities which can be derived from $\rsqL$ such as
the gyration radius~\cite{RosaEveraersPRL2014}, or the bond-vector orientation correlation function, Eq.~(\ref{eq:bacf}) in the main text, which
decays on the Kuhn scale and drops to $-l_K/L$ for large distances as a consequence of the closure constraint (Fig.~S\ref{fig:Klein}b).
The difference between the random walk ribbons and Gaussian rings only becomes apparent from
the asymmetry ratios of the gyration tensor~\cite{RosaEveraersPRL2014}:
by construction, for large Klein ribbons we find the typical values for ordinary random walks $\approx 11.7:2.7:1.0$~\cite{rudnick},
at odds with the measured $\approx 6.1:2.3:1.0$ for Gaussian rings~\cite{bishop_michels_jcp1985}.

By construction, the linear ribbon model predicts the $L^{1/2}$ growth of the overlap parameter, $\Omega(L)\equiv \frac{\rho_K l_K}{L} { \langle R^2 (L) \rangle }^{3/2}$,
which is characteristic for linear chains (Fig.~S\ref{fig:Klein}d).
For ring sizes up to a few entanglement lengths,
long (up to $\approx 10^6 \tau_e$, see Table~S\ref{tab:MDruns}) MD equilibration runs of Klein folded initial states hardly affect the conformational statistics.
However, larger rings undergo substantial shrinking (Fig.~S\ref{fig:Klein}a) with correspondingly increased contact probabilities
(Fig.~S\ref{fig:Klein}c),
develop anti-correlations in the bond-vector orientation correlation function on the entanglement scale (Fig.~S\ref{fig:Klein}b),
and lower the overlap parameter (symbols in Fig.~S\ref{fig:Klein}d) slightly below
the entanglement threshold, $\Omega \equiv 20$~\cite{KavassalisNoolandiPRL1987,FettersMacromolecules1994,uchida}.

\begin{figure}
\includegraphics[width=0.48\textwidth]{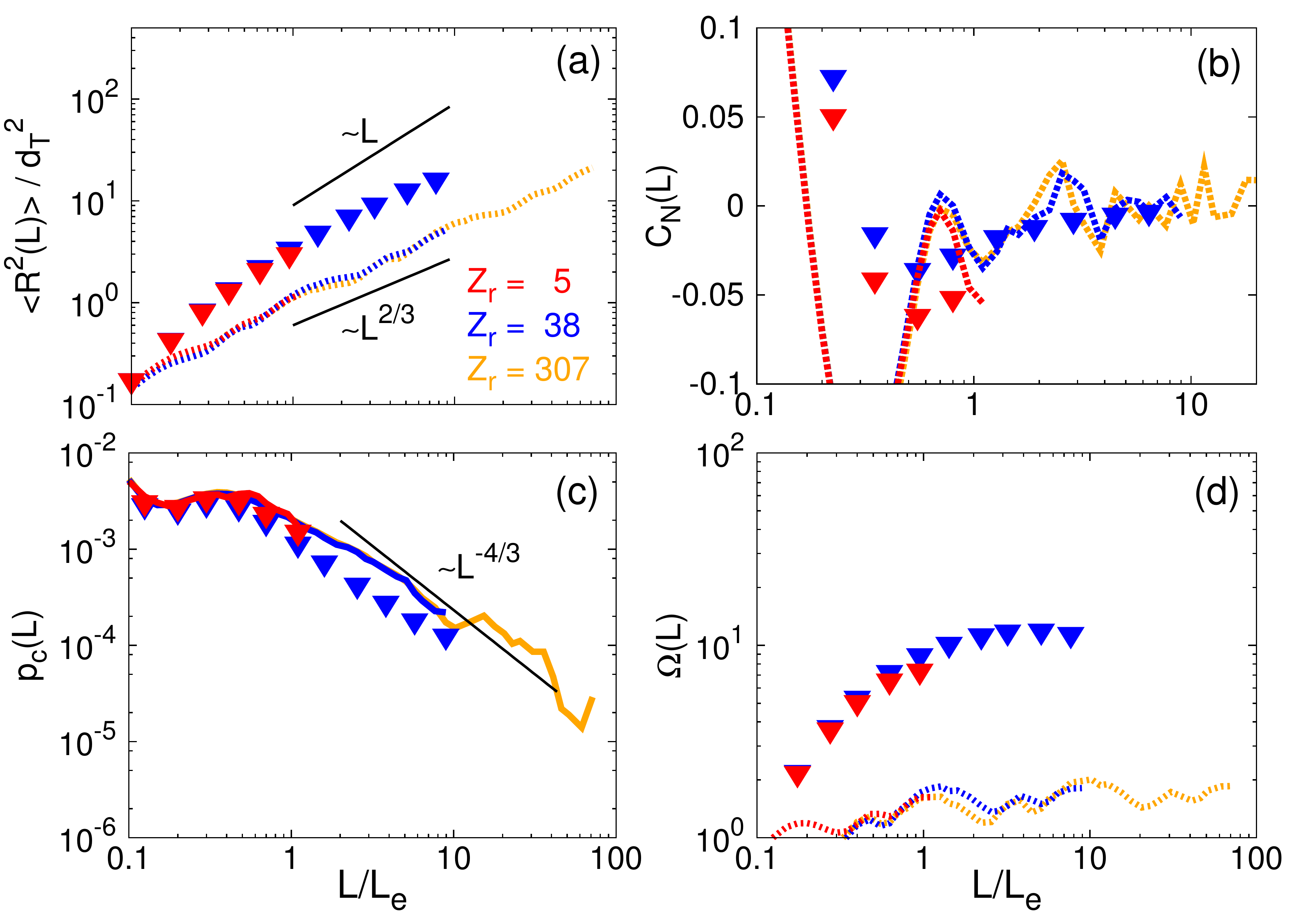}
\caption{
\label{fig:Moore}
Comparison of the conformational statistics for MD-equilibrated rings (symbols) and space-filling Moore rings
(lines:
dotted, average at $t=0$;
dashed, average at $t=0.1\tau_e$;
solid, average at $t=\tau_e$).
}
\end{figure}

{\bf Moore rings} --
In melts derived from standard space-filling curves neighboring rings do not overlap at all.
The lines in Fig.~S\ref{fig:Moore} represent the conformational properties of Moore rings.
Panel (a) and panel (c) (whose curves were averaged after a short MD run up to $\tau_e$)
show that $\rsqL\sim L^{2/3}$ and $p_c(L)\sim L^{-4/3}$ in agreement with~\cite{hic}.
The regular structure manifests itself in an oscillating bond-vector orientation correlation function
(Fig.~S\ref{fig:Moore}b, curves averaged after a short MD run up to $\tau_e/10$,
sufficient to equilibrate the chain statistics below the entanglement scale).
Interestingly, the overlap parameter of $\approx 2$~\cite{Moore_note} never approaches the entanglement threshold of $\Omega = 20$ (Fig.~S\ref{fig:Moore}d).

We have also performed long (up to $\approx 10^5 \tau_e$, see Table~S\ref{tab:MDruns}) MD simulations to equilibrate systems with $Z_r=5$ and $Z_r=38$ (symbols in Fig.~S\ref{fig:Moore}).
In our final conformations the oscillations in the bonds orientations are again replaced by anti-correlations on the entanglement scale.
In particular, the Moore conformations undergo substantial swelling,
increasing the overlap parameter on large length scales close to the entanglement threshold.

\begin{figure}
\includegraphics[width=0.48\textwidth]{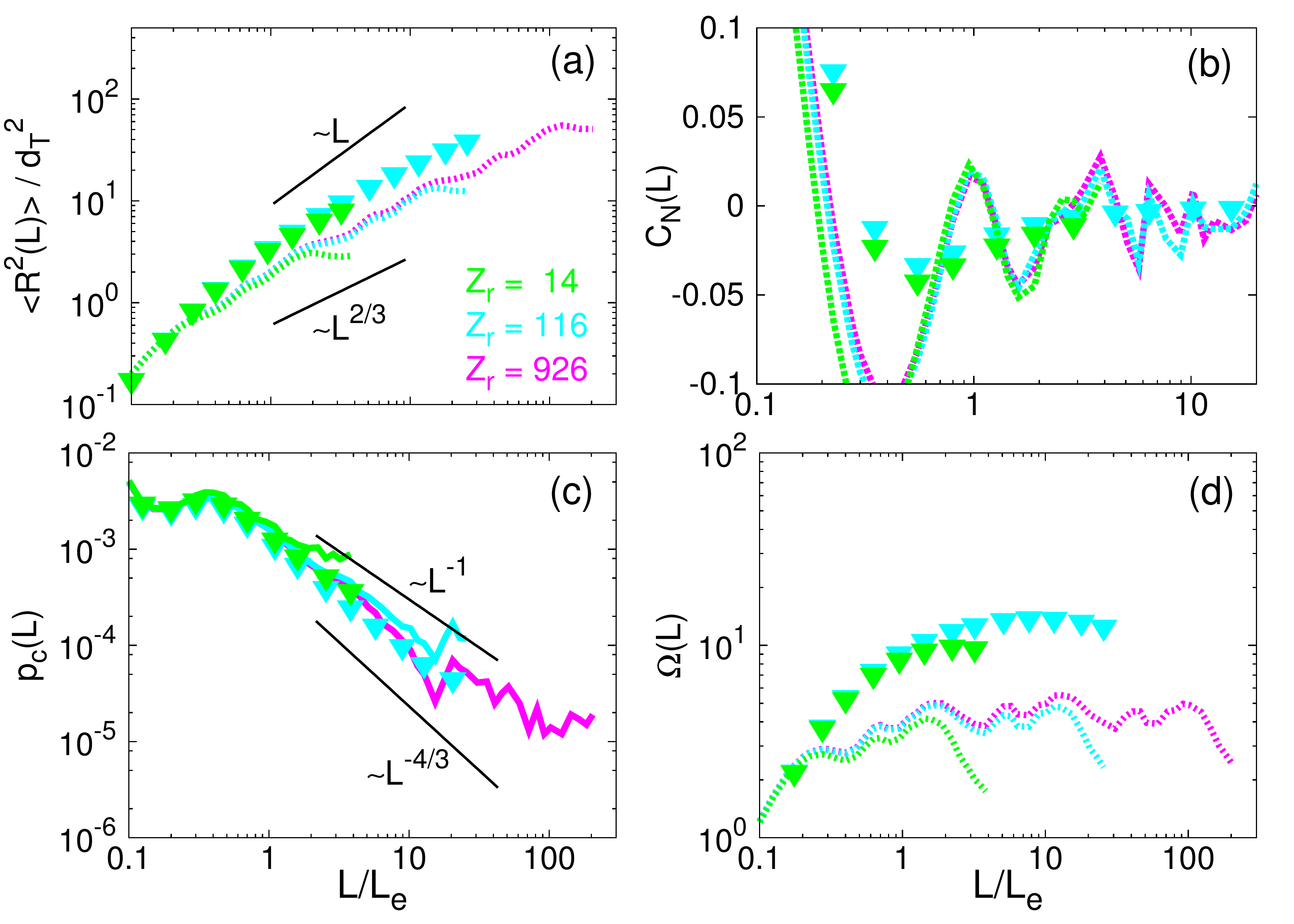}
\caption{
\label{fig:Hilbert}
Comparison of the conformational statistics for MD-equilibrated rings (symbols) and space-filling Hilbert ribbons
(lines:
dotted, average at $t=0$;
dashed, average at $t=0.1\tau_e$;
solid, average at $t=\tau_e$).
}
\end{figure}

{\bf Hilbert ribbons} --
The Hilbert ribbons have a similar conformational statistics as Moore rings (Fig.~S\ref{fig:Hilbert}).
The typical size grows like $\rsqL\sim L^{2/3}$ as long as $L\ll L_r$.
The conformations are locally less crumpled.
The overlap parameter (Fig.~S\ref{fig:Hilbert}d) of $\approx 5$ is nearly twice as large as for Moore rings~\cite{HilbertRibbons_note},
but it stays nevertheless well below the entanglement threshold.
Interestingly, contact probabilities decay like $p_c(L)\sim L^{-1}$ (Fig.~S\ref{fig:Hilbert}c) in better agreement with the experimental~\cite{hic} and simulation data~\cite{RosaBJ2010}.
For $Z_r = 14$ and $Z_r = 116$ we have prepared (see Table~S\ref{tab:MDruns}) equilibrated melt conformations starting from $M=8$ chains with identical Hilbert ribbon conformations (symbols in Fig.~S\ref{fig:Hilbert}).
Again, the oscillations in the bonds orientations are replaced by anti-correlations on the entanglement scale with the rings swelling close to the entanglement threshold.

\begin{figure}
\includegraphics[width=0.48\textwidth]{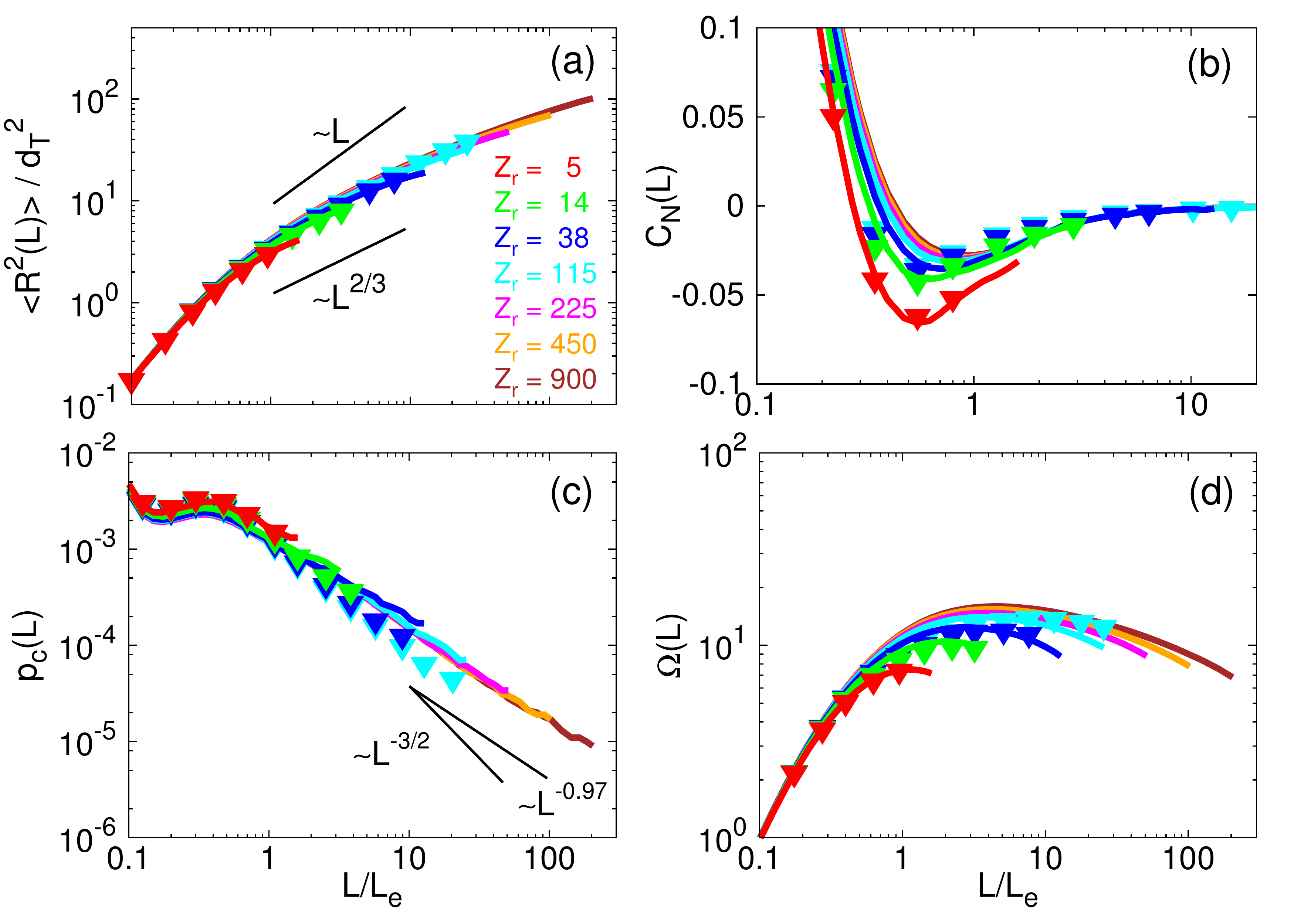}
\caption{
\label{fig:IdealLatticeAnimal}
Comparison of the conformational statistics for MD-equilibrated rings (symbols) and ideal lattice tree ribbons
(solid lines, average at $t=\tau_e$).
Notice, that the large-scale decay of contact probabilities, $p_c(L) \sim L^{-0.97 \pm 0.01}$ (panel (c)), is different from the observed behavior
of rings obtained from the interacting lattice tree model (see panel (c1) of Fig.~\ref{fig:internal ring statistics} in the main text).
Note also the slow divergence of the overlap parameter with chain length (panel (d)).
}
\end{figure}

{\bf Branched ribbon conformations from the {\it ideal} lattice tree model} --
Results for rings derived from melts of ideal lattice trees are illustrated in Fig.~S\ref{fig:IdealLatticeAnimal} (solid lines).
We note the characteristic anti-correlations of bonds orientations (panel (b)) on the entanglement scale and the overlap parameter just below the entanglement threshold (panel (d)).
Again, we have constructed topologically correct melt states by assembling single ring conformations at the correct monomer
density into a simulation box with periodic boundary conditions.
Starting from these, we have run MD simulations for as long as in the previous cases
({\it i.e.} $\approx 10^5 \tau_e$, see Table~S\ref{tab:MDruns}).
Results for MD simulations are summarized as symbols in Fig.~S\ref{fig:IdealLatticeAnimal}.
We notice that, while small rings (up to $Z_r=38$, blue symbols) are well described by the ideal lattice tree model,
the cyan system ($Z_r=115$) starts showing some swelling, especially evident in the overlap parameter (panel (d)).

\section{Brute-force Molecular Dynamics (MD) equilibration of ``gold standard'' reference data}\label{sec:Brute force equilibration}

Table~S\ref{tab:MDruns} lists the specifications of the ring melts, which we have studied~\cite{RosaEveraersPRL2014} using Molecular Dynamics simulations. 

\begin{figure}
\includegraphics[width=3.3in]{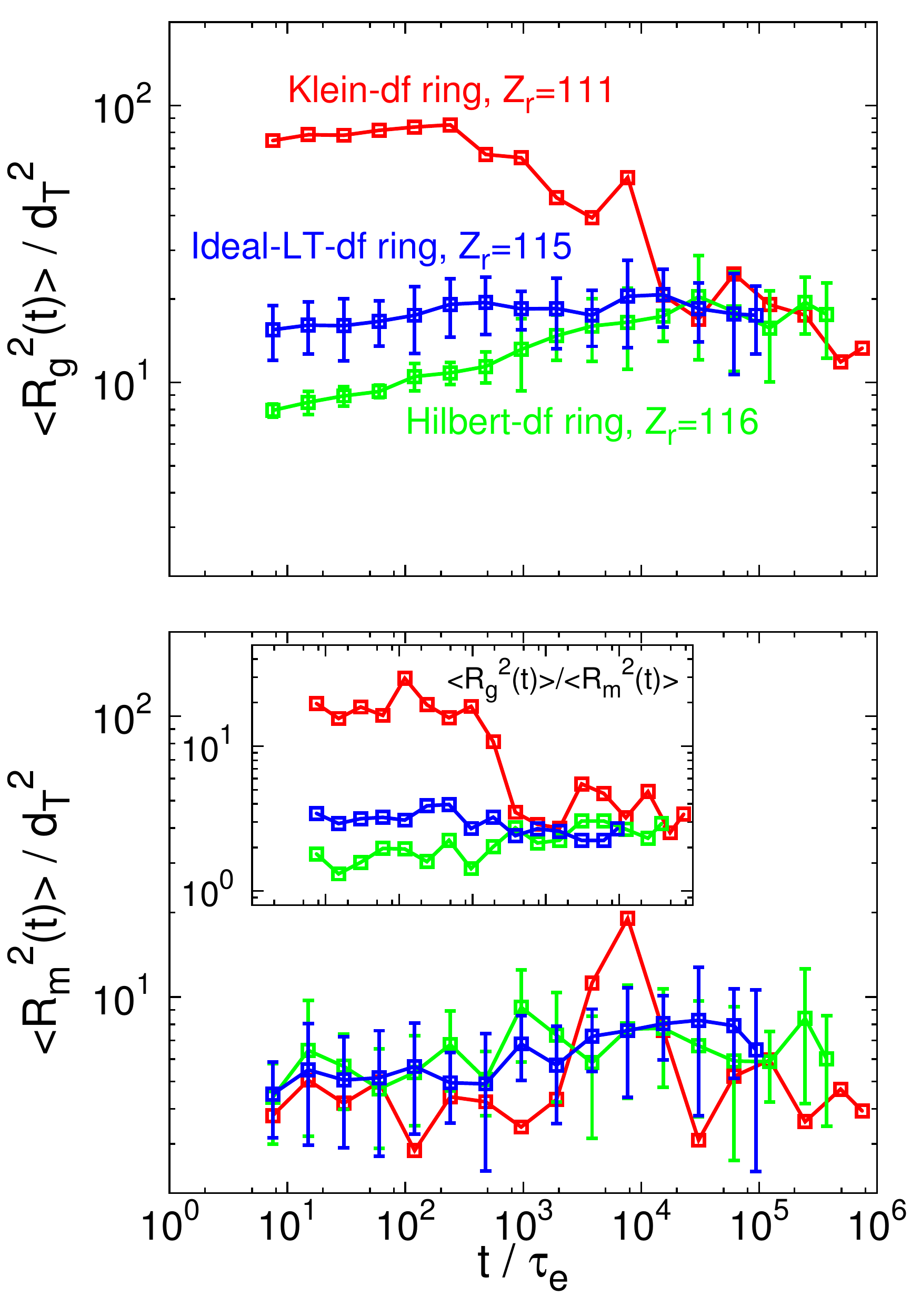}
\caption{
\label{fig:KleinHilbertRelaxation}
Time behaviors of
the average square gyration radius ($\langle R_g^2(t) \rangle $, top)
and
the average square magnetic radius ($\langle R_m^2(t) \rangle$, bottom) 
for solutions of ring polymers with different initial conformations:
(1)
Klein double-folded rings (red),
(2)
Hilbert double-folded rings (green)
and
(3)
double-folded rings on ideal branched primitive paths (blue).
Systems (2) and (3) are made of 8 rings, while system (1) is made of only one chain.
The inset shows the ratio $\langle R_g^2(t) \rangle / \langle R_m^2(t) \rangle $.
Error bars are for the standard deviation of the mean.
}
\end{figure}
\begin{figure}
\includegraphics[width=3.3in]{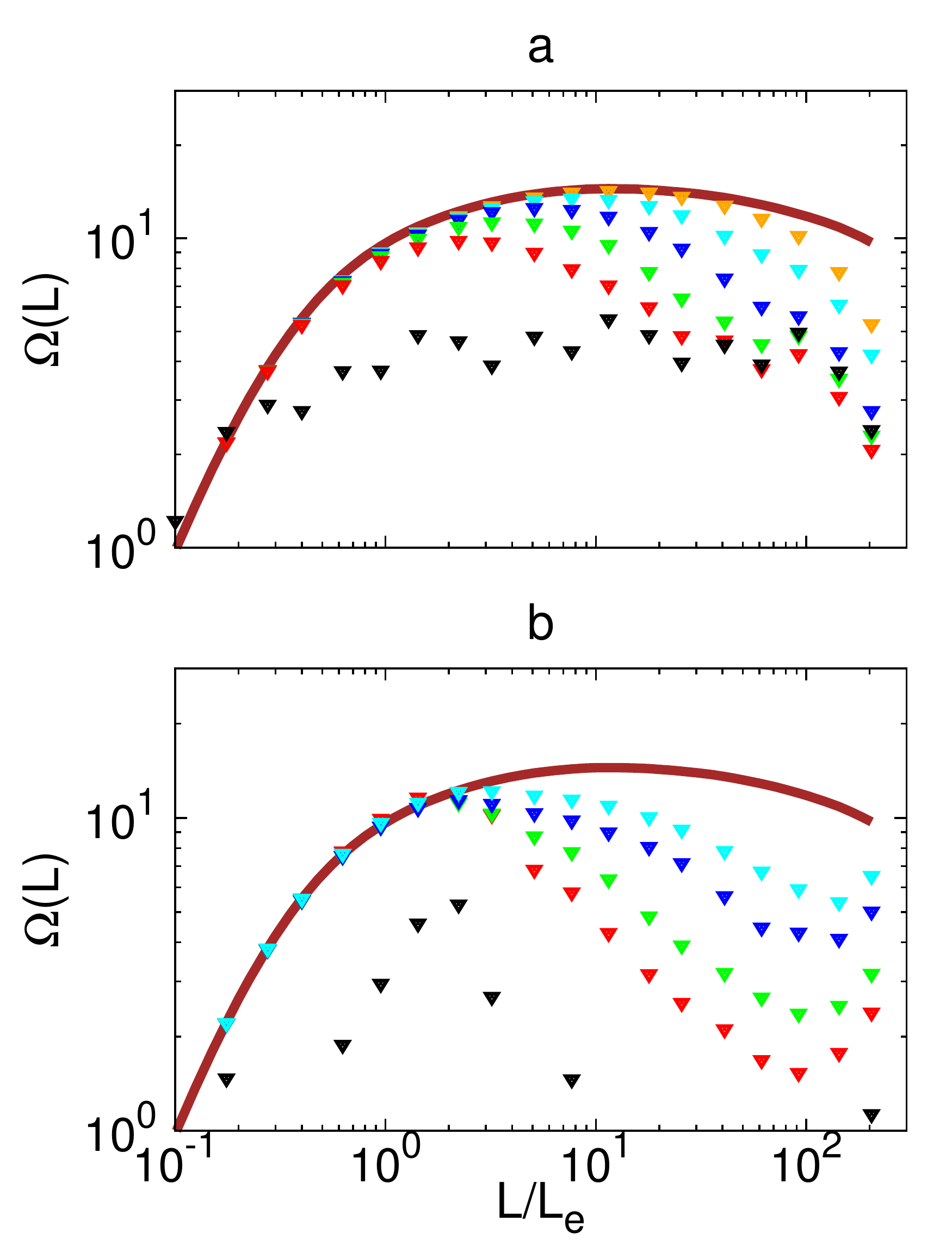}
\caption{
\label{fig:MDrelaxation}
MD-equilibration of overlap parameter, $\Omega(L)$, shows the clear progression from small to large contour-length separations.
Symbols of different colors correspond to square internal distances averaged over exponentially larger and larger time windows:
black symbols show data corresponding to the initial configuration;
symbols from red to orange represent MD data averaged over $10^i < t/\tau_e < 10^{i+1}$, with $i$ from $0$ to $4$ respectively
(there are no data for $i=4$ in panel (b)).
The two panels show:
(a)
the largest ($Z_r=926$) Hilbert ribbon (see Table~S\ref{tab:MDruns} here),
(b)
the linear polymer chains the size the human chromosomes ($Z=810$) studied in~\cite{RosaPLOS2008}.
The brown solid line is the prediction of the interacting lattice tree model.
}
\end{figure}
\begin{figure*}
\includegraphics[width=1.0\textwidth]{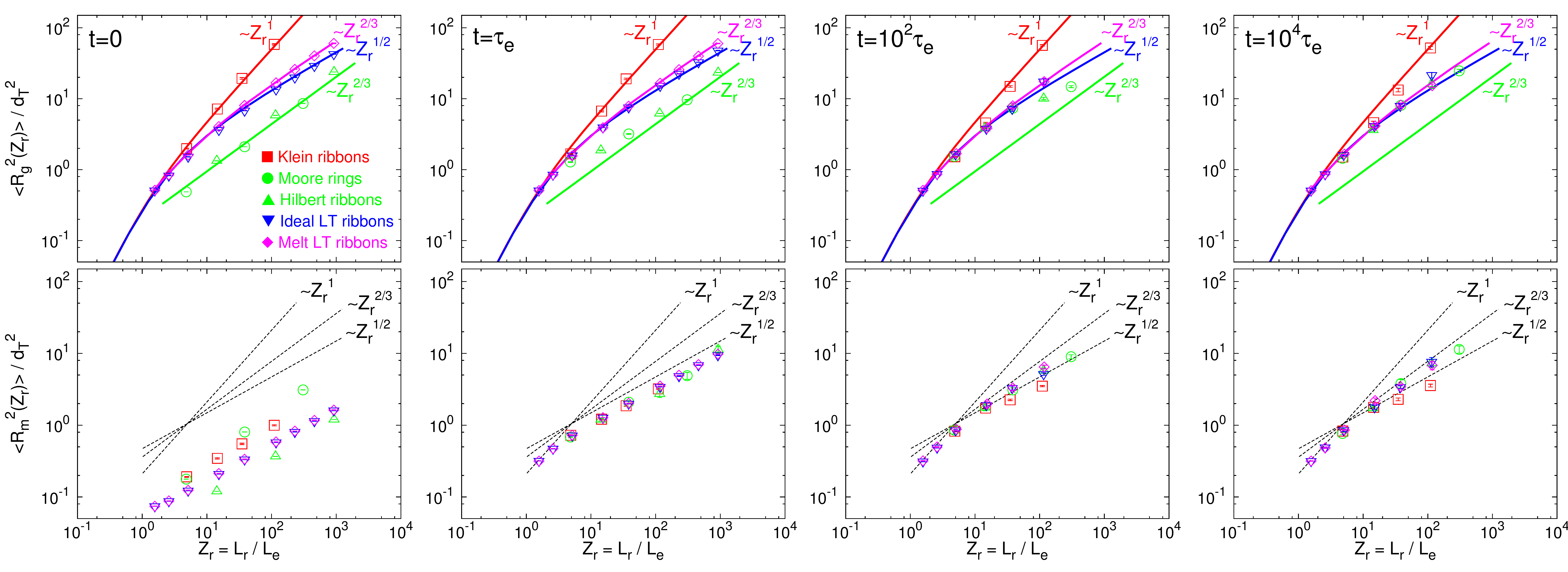}
\caption{
\label{fig:Rg2andRm2Equilibration}
Equilibration of gyration (top row) and magnetic (bottom row) radii in Molecular Dynamics simulations of the fiber model.
Lines with different colors correspond to the theoretically expected gyration radii for the different types of structures.
Symbols refer to simulation data for ensembles derived from corresponding initial states.
Not all systems were run to the maximal time (see Table~S\ref{tab:MDruns} here), explaining the different number of symbols used in the different panels. 
}
\end{figure*}

Fig.~S\ref{fig:KleinHilbertRelaxation} illustrates the time evolution for the average gyration and magnetic radii ($\langle R_g^2(t) \rangle$ and $\langle R_m^2(t) \rangle$, top and bottom panel respectively)
for our largest ring polymers with $Z_r\approx100$ and different initial conformations (see Sec.~\ref{sec:Lattice model of double folded rings} in the main text, Sec.~\ref{sec:SM Fractals} here and Table~S\ref{tab:MDruns}).
As expected, after a long transient the memory of the initial conformation is lost and both observables fluctuate around their corresponding equilibrium values with ratio $\langle R_g^2(t) \rangle / \langle R_m^2(t) \rangle \approx 2.1$ (see inset of bottom panel).
As in the case of relaxation of long, untangled linear chains~\cite{RosaPLOS2008}, rings equilibration proceeds from small to large scales (Fig.~S\ref{fig:MDrelaxation}).
As illustrated by Fig.~S\ref{fig:Rg2andRm2Equilibration}, the magnetic and the gyration radius equilibrate on similar time scales. 
In particular, this implies that the lattice tree melt derived ring melts are {\it not} equilibrated and would need to be brute-force equilibrated like any other topologically correct initial state.

\end{document}